\documentclass{cernyrep} 
\usepackage{texnames}
\usepackage[T1]{fontenc}

\usepackage{epsfig}
\usepackage{amsmath}
\usepackage{amssymb}
\newcommand{\Alpgen}{{\tt Alpgen}{}}
\newcommand{\Madgraph}{{\tt Madgraph}{}}
\newcommand{\Sherpa}{{\tt Sherpa}{}}
\newcommand{\MCFM}{{\tt MCFM}{}}
\newcommand{\ggtoWW}{{\tt gg2WW}{}}
\newcommand{\ggtoZZ}{{\tt gg2ZZ}{}}
\newcommand{\VBFNLO}{{\tt VBFNLO}{}}
\newcommand{\GoSam}{{\tt GoSam}{}}

\newcommand{\HELACNLO}{{\tt HELAC-NLO}{}}
\newcommand{\HELAConeLOOP}{{\tt HELAC-1LOOP}{}}

\newcommand{\Pythia}{{\tt Pythia}{}}
\newcommand{\Herwig}{{\tt Herwig}{}}
\newcommand{\Herwigpp}{{\tt Herwig++}{}}
\newcommand{\POWHEG}{{\tt POWHEG}{}}
\newcommand{\POWHEGBOX}{{\tt POWHEG BOX}{}}
\newcommand{\MCatNLO}{{\tt MC@NLO}{}}

\newcommand\POWHEGpPYTHIA{{\tt POWHEG+PYTHIA}}

\newcommand{\MENLOPS}{{\tt MENLOPS}{}}
\newcommand{\LoopSim}{{\tt LoopSim}{}}

\begin{document}
\title{QCD and  Collider Processes}
 
\author{Giulia Zanderighi}

\institute{University of Oxford and STFC, 
        Theoretical Physics, 1 Keble Road, OX1 3PN Oxford, UK}

\maketitle % this produces the title block

 \begin{abstract}
  I review the status of today's theoretical description of Standard
  Model processes relevant for Tevatron and LHC analyses, and of the
  tools that are used in phenomenological studies. I will also discuss
  a few recent ideas to further refine our abilities to perform
  technically challenging calculations.
\end{abstract}

\section{Introduction}
Today we face many fundamental questions, some of which are driven by
experimental data, such as the question about the mechanism of
electroweak (EW) symmetry breaking, the nature of dark matter, and the
physics associated with the vacuum energy, as well as questions that
are driven by theoretical curiosity and ambition, which are in essence
propelled by our hope that there is an elegant structure behind what
we observe in nature. Questions of the latter type include why there
are three generations, what causes the hierarchy of fermion masses and
mixing, and how the strong CP problem is resolved.
While questions of the first type do have a definite answer, this is
not necessarily the case for the questions of the second type. From
this point of view, some of these questions might not even be the
right ones to ask. With the start of the LHC we feel that we are at a
verge of big changes, the depth of which we can not assess
yet. Indeed, the LHC marks the start of a long research program and
experiments at the LHC are expected to revolutionize our understanding
of the fundamental forces and matter. The LHC will definitely explore
the origin of mass and the associated nature of EW symmetry breaking.
In the course of this, it might also shed light on the nature of dark
matter and the origin of the matter-antimatter asymmetry. It may also
explore the physics that underlies the evolution of the early
universe.
While it is clear that the LHC will not answer all the fundamental
questions that we have, the questions we ask now will most likely
change after the LHC era.  It is therefore really a great and unique
time to be a particle physicist. 

Since the beginning of its run in 2010, the LHC has been remarkable
successful.
ATLAS and CMS collected around 45 pb$^{-1}$ in 2010, more than 1
fb$^{-1}$ by the time of summer conferences, and more than 5 fb$^{-1}$
by the end of the year. At the end of the 2011 run, almost every week
has set a new record in instantaneous luminosity. With the 2010 and
early 2011 data, remarkably, all major Standard Model (SM) processes
have already been re-established, including single-top and di-boson
production, challenging measurements (because of the small/large
cross-sections/backgrounds) that have been performed at the Tevatron
only in recent years. By now, we have entered a new territory in the
search of physics beyond the SM (BSM) with sensitivities already well
exceeding those of LEP and the Tevatron.

One important question then concerns the role of QCD for LHC
measurements and new-physics searches.  Understanding how QCD works is
essential in order to make accurate predictions for both the signal
and background processes. This typically requires complex calculations
to higher orders in the perturbative expansion of the coupling
constant. Understanding QCD dynamics can however also help reduce
backgrounds and sharpen the structure of the signal.  This can for
instance be achieved by designing better observables, by employing
appropriate jet algorithms, by using jet-substructure, or by
exploiting properties of boosted kinematics. Finally, once discovery
is made, QCD will be crucial to extract the properties (masses, spins,
and couplings) of the new states found. Therefore, at the LHC, no
matter what physics you do, QCD will be part of your life.

\begin{figure}[thpb]
  \centering
\vspace{0.15cm}
\includegraphics[angle=90,scale=0.35]{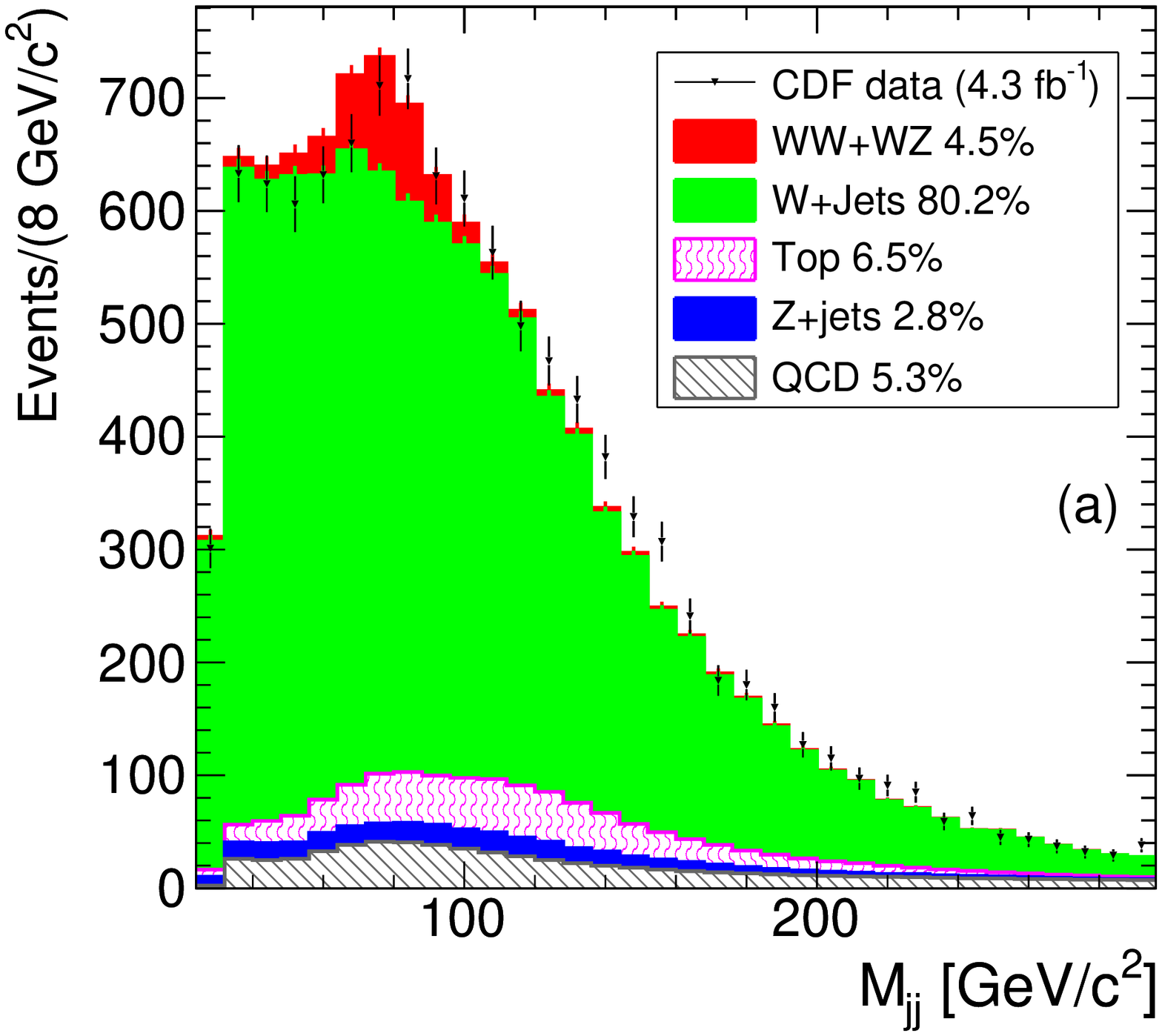}
\hspace{1cm}
\includegraphics[angle=90,scale=0.35]{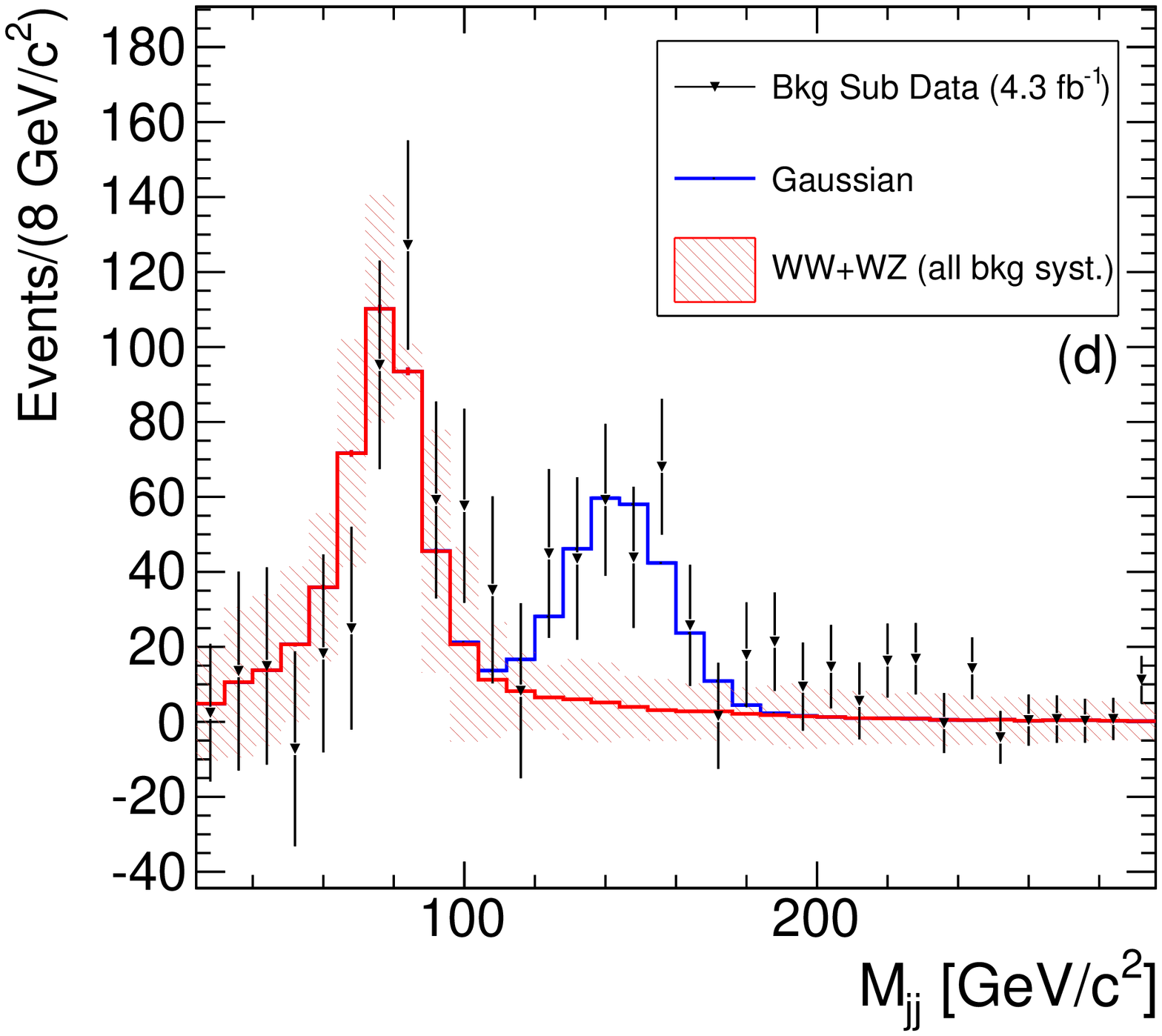}
\caption{The dijet invariant mass distribution for $W$ plus dijet
  events as measured by CDF. The left pane shows the fits for known
  processes only. The right pane shows, by subtraction, the resonant
  contribution to $m_{jj}$ including $WW$ and $WZ$ production and a
  hypothesized narrow Gaussian contribution. Figures taken
  from~\cite{Aaltonen:2011mk}.}
\label{fig:Wjj}
\end{figure}

It is interesting to first recall a recent measurement that was the
origin of considerable excitement. In April 2011, CDF reported the
observation of a peak in the $m_{jj}$ distribution in $W$ + dijet
events~\cite{Aaltonen:2011mk}, see Fig~\ref{fig:Wjj}. The first
measurement had a $3.2\sigma$ significance, and was based on $4.3$
fb$^{-1}$. Subsequently, more data ($7.3$ fb$^{-1}$) has been
analyzed, leading to a significance of more than
$4\sigma$~\cite{cdfwebpage}.  Since then, a large number of tentative
new-physics explanations appeared on the arXiv, along with a few SM
analyses that address the question of whether this effect can be
attributed to a mismodelling of one of the SM backgrounds (in
particular single
top)~\cite{Plehn:2011nx,Campbell:2011gp,Sullivan:2011tf}. The
excitement was curbed few months afterwards, when in June, D0
announced that it did not confirm the excess seen by
CDF~\cite{Abazov:2011af}. It is yet unclear what the reasons for the
discrepancy between CDF and D0 findings are, if any. However, this
example demonstrates that even in the case where one identifies a mass
peak in the tail of a distribution (a scenario that was considered
``an easy discovery'') a robust control of SM backgrounds remains
mandatory, in particular when the shape of the backgrounds is one of
the issues. Currently we have a number of other recent measurements at
collider experiments that report a few sigma deviations from the SM
predictions. This is for instance the case for the top
forward-backward asymmetry measured by both CDF~\cite{Aaltonen:2011kc}
and D0~\cite{Abazov:2011rq}, for the dimuon charge asymmetry measured
by D0~\cite{Abazov:2011yk}, for $W+b$ measured by
D0~\cite{Aaltonen:2009qi}, CP violation in time-integrated $D^0$
decays~\cite{Aaij:2011in}, and a few more.

The important question becomes then what the tools at our disposal are
to make precise predictions, and whether we have the solid control of
signal and background processes that is needed in order to claim
discoveries.
In this write-up I will discuss the status of our theoretical
knowledge of most important SM processes, the tools at our disposal to
describe these processes, and the impact of QCD higher orders in view
of recent Tevatron/LHC results.  Finally, I will discuss a few recent
ideas to further improve on the way we perform technically challenging
calculations.

\section{Perturbative tools}
The range of physics analyses that one can do at the LHC is very
broad. It includes pure instrumental QCD studies, such as measurements
of parton densities and inclusive jet cross-section measurements,
precision EW measurements, Higgs searches, direct and indirect BSM
searches, $B$ physics, top physics, diffractive studies and forward
physics, and heavy ion physics. Each of these topics includes a vast
number of measurements and studies. Yet, there are three things that
everybody involved in any of these analyses can not live
without. These are Monte Carlos (MCs), parton distribution functions
(PDFs), and jets.

\subsection{Monte Carlos and leading order matrix elements}
The first thing one can not live without at the LHC are MC generators.
Apart from very few exceptions, every analysis at the LHC uses a MC
program for the simulation of the signal process, for the backgrounds,
for subtracting the underlying event and the non-perturbative
contributions, and/or for efficiency studies and modeling of the
detector response. The current level of sophistication is such that
essentially not a single study relies on \Pythia/\Herwig{} alone. It
is well understood that in multi-parton processes it is important to
describe the multiple hard QCD radiation at least using exact matrix
elements, employing for instance \Alpgen~\cite{Mangano:2002ea},
\Madgraph~\cite{Alwall:2011uj}, or \Sherpa~\cite{Gleisberg:2008ta}.

\begin{figure}[t]
  \centering
\includegraphics[angle=0,scale=0.30]{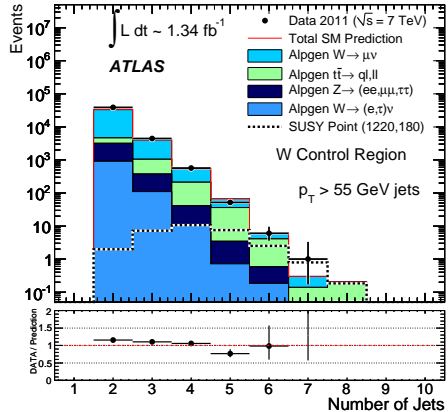}
\caption{
The multiplicity of jets with $p_T$ > 55 GeV for events in a $W$
enriched control region at the LHC (7 TeV). Figure taken
from~\cite{Aad:2011qa}.} 
\label{fig:6j}
\end{figure}

Since experimental studies rely heavily on all these leading-order
(LO) tools, there is continuous progress in their development, and the
\Herwig/\Pythia{} codes that we have today bear little resemblance to
their original version of the '80s.
In particular, in \Pythia{} 8.1~\cite{Sjostrand:2007gs} (a C++ code)
there is a new fully interleaved $p_{\rm t}$-ordered multi-parton
interaction (MPI), initial- and final state evolution (the original
mass-ordered evolution is not supported any longer), a richer mix of
underlying event processes ($\gamma$, $J/\Psi$, DY), the possibility
to select two hard interactions in the same event, an $x$-dependent
proton size in the MPI framework, the full hadron--hadron machinery
for diffractive systems, several new processes in and beyond the SM,
and various other new features.
\Herwigpp~\cite{Gieseke:2011na} (the current version is 2.5.2) has new
next-to-leading order (NLO) matrix elements, including weak boson pair
production, a colour reconnection model, diffractive processes,
additional models of BSM physics, and new LO elements for
hadron-hadron, lepton-lepton collisions, and photon-initiated
processes.
\Sherpa~\cite{Gleisberg:2008ta} (version 1.3.1) has improved integration
routines in Comix, a simplified kinematics reconstruction algorithm of
the parton shower (PS), leading to numerically more stable
simulations, HepMC output for NLO events and various other
improvements/bug-fixes.
\Madgraph~\cite{Alwall:2011uj} (version 5) has a completely new
diagram generation algorithm, which makes optimal use of
model-independent information, has an efficient decay-chain package,
and a new library for the colour calculations. Some applications from
{\tt ALOHA}{}~\cite{deAquino:2011ub}, an automatic library of helicity
amplitudes for Feynamn diagram computations are also implemented.
Altogether, there is continuous, fast progress in various directions.
So far, it is amazing how well these tools work, once the
normalization is fixed using data. A very recent comparison of data
with \Alpgen{} up to six jets (a control region for BSM searches) is
shown as an illustration in Fig.~\ref{fig:6j}~\cite{Aad:2011qa}.  Yet,
the devil is often in the detail (i.e. in the $\sim 20-30$\% effects
that are opposite to expectations). For instance in general one
expects matrix-element based MCs to work better than pure PSs, but
this is not always the case (see e.g. ~\cite{Khachatryan:2011dx}).
Altogether, these LO programs will undergo a real stress test in the
coming years.

\subsection{The NLO revolution}
\label{sec:NLO}
Theorists like to advertise NLO computations by using the reduction of
scale uncertainties in the predictions as an argument, which is meant
to reflect the reduction in the theoretical perturbative
uncertainty. However, the strongest argument in support of NLO
calculations is their past success in accurately describing LEP and
Tevatron data. Because of the importance of NLO corrections, an
industrial effort has been devoted in the last years to these
computations~\cite{Binoth:2010ra}.
Recent revolutionary ideas in the way NLO computations are
performed include
sewing together tree-level amplitudes to compute loop amplitudes
(using on-shell intermediate states, cuts, unitarity ideas,
\ldots)~\cite{Britto:2004nc},
the OPP algorithm, an algebraic way to extract coefficients of
master integrals by evaluating the amplitudes at specific values of
the loop momentum~\cite{Ossola:2006us},
and $D$-dimensional unitarity, a practical numerical tool to
evaluate full amplitudes, including the rational part, with unitarity
ideas~\cite{Giele:2008ve}.
For a pedagogical review on unitarity methods see~\cite{Ellis:2011cr}.
These methods led in the past 2 to 3 years to a number of 2 $\to$ 4
calculations at hadron colliders. These include $W+3$
jets~\cite{KeithEllis:2009bu,Berger:2009zg}, $Z+3$
jets~\cite{Berger:2010vm}, $t \bar t b \bar
b$~\cite{Bevilacqua:2009zn}, $t \bar t \to W^+W^- b \bar
b$~\cite{Bevilacqua:2010qb}, $W^+W^+$ + 2 jets~\cite{Melia:2010bm},
$W^+W^-$ + 2 jets~\cite{Melia:2011dw}, $ b \bar b b \bar
b$~\cite{Greiner:2011mp}, $t \bar t $ + 2
jets~\cite{Bevilacqua:2011hy}, four jet production~\cite{Bern:2011ep},
and a few other ones.

Feynman diagram methods have also been applied successfully to 2 $ \to
$ 4 calculations, this is for instance the case for quark-induced $b
\bar b b \bar b$~\cite{Binoth:2009rv}, $t\bar tb\bar
b$~\cite{Bredenstein:2010rs}, $W^+W^-b\bar b$~\cite{Denner:2010jp}
production, $W\gamma\gamma$+jet~\cite{Campanario:2011ud}, and a number
of vector boson fusioon (VBF) processes which are available in the
public code \VBFNLO~\cite{Arnold:2011wj}. VBF processes have never
been measured at the Tevatron. First measurements of VBF $W$ or $Z$
production at the LHC are therefore of particular interest. They also
pave the way to measurements of VBF Higgs production.
Note that only a few years ago, performing this type of calculation
with Feynman diagrams was considered an impossible task.

A novel approach to NLO calculations, based on the results
of~\cite{vanHameren:2009vq}, promotes traditional tree algorithms to
generators of loop-momentum polynomials, that are called open
loops~\cite{Cascioli:2011va}. The excellent performance of the method
has been demonstrated, open loops have therefore a potential to
address a number of multi-particle processes at hadron colliders.
Another recent approach evaluates one-loop QCD amplitudes purely
numerically~\cite{Becker:2010ng}. The algorithm consists of
subtraction terms, approximating the soft, collinear and ultraviolet
divergences of one-loop amplitudes and a method to deform the
integration contour for the loop integration into the complex
space. The algorithm is formulated at the amplitude level and does not
rely on Feynman graphs. The power of this method has been demonstrated
recently with the leading color NLO calculation of up to seven jets in
$e^+e^-$ collisions~\cite{Becker:2011vg}.

\begin{figure}[t]
  \centering
  \includegraphics[clip,scale=0.5]{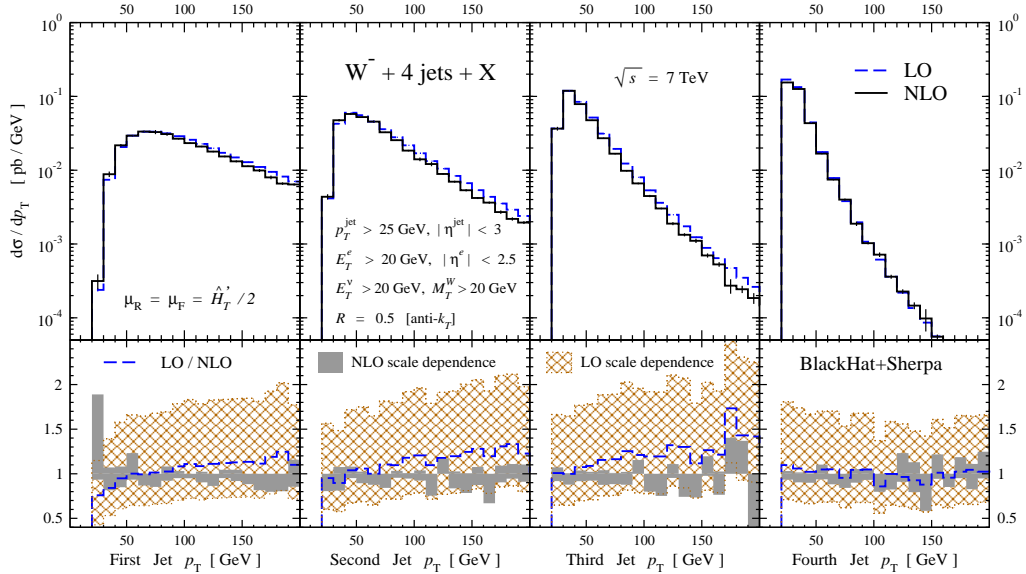}
  \caption{The transverse momentum distribution of the leading four
    jets in $W^- + 4$-jet production at the LHC (7 TeV) at LO and
    NLO. The lower panels show the LO and NLO scale-dependence bands
    ($\hat H'/4\le \mu_R=\mu_F\le \hat H'_T$) normalized to the
    central NLO prediction ($\mu_R=\mu_F=\hat H'_T/2$). Figure taken
    from~\cite{Berger:2010zx}.}
\label{fig:W4j}
\end{figure}

Given that both Feynman diagram and unitarity based methods allowed us
to compute $2 \to 4$ processes at NLO in QCD, it might be unclear
where the revolution advocated in the heading of the subsection lies
in. The revolution, I believe, is not yet in the applications that we
see today, rather in the prospect for low-cost fully
computer-automated NLO calculations even beyond $2 \to 4$ in the near
future.  Indeed, two $2\to 5$ processes have already been computed at
NLO, namely $W$ + 4 jets~\cite{Berger:2010zx} and $Z$ + 4
jets~\cite{Ita:2011wn}.\footnote{In both cases the leading colour
  approximation has been used. This approximation is expected to give
  rise to very small (percent) corrections only.}  Fig.~\ref{fig:W4j}
illustrates in the case of $W^-+4$ jets the typical effect of
including NLO corrections: one obtains a considerable reduction of the
scale uncertainty, and, for some distributions, a change in shape.
As far as the full automation is concerned, let me highlight three
interesting approaches. The first one~\cite{Hirschi:2011pa} is a
method based on Feynman diagrams, it uses the OPP procedure for the
virtual calculation, and the FKS subtraction of divergences, together
with clever and efficient procedures to deal with instabilities. More
improvements and refinements are to be expected soon. At present there
is no public code, instead the idea is to provide $N$-tuples.  The
second approach, \HELAConeLOOP~\cite{Bevilacqua:2011xh}, is a program
that evaluates numerically QCD virtual corrections to scattering
amplitudes using the OPP method. The public program is part of the
\HELACNLO{} framework that allows for a complete evaluation of QCD NLO
corrections.
\GoSam~\cite{Cullen:2011ac} is a third approach which aims at the full
automated calculation of NLO corrections for multi-particle
processes. The one-loop amplitudes are generated using Feynman
diagrams and are reduced using D-dimensional unitarity, a refined
Passarino-Veltman style tensor reduction, or a combination of
both. \GoSam{} can be used to calculate one-loop corrections to both
QCD and electroweak theory, and model files for BSM theories can be
also linked. An interface to programs calculating real radiation is
included too. The flexibility of \GoSam{} has been demonstrated
explicitly by considering various examples.

\subsection{Merging NLO and Parton Showers}
\label{sec:NLOplusPS}
While NLO predictions provide relatively accurate results for
inclusive cross-sections, they do not furnish an exclusive description
of the final state that can be compared with actual particles in the
detectors, as MC programs do. It is therefore useful to combine the
best features of both approaches. Two public frameworks exist for this
purpose, namely \MCatNLO~\cite{Frixione:2010wd} and
\POWHEG~\cite{Nason:2004rx}. These tools are almost 10 years old now,
and since their conception a long list of processes has been
implemented in both frameworks.
In particular, recently the \POWHEGBOX{} was
released~\cite{Alioli:2010xd}, which is a general framework for
implementing NLO calculations in shower MC programs according to the
\POWHEG{} method.  The user only needs to provide a simple set of
routines (Born, colour-correlated Born, virtual, real, and phase
space) that are part of any NLO calculation.

\begin{figure}[t]
  \centering
\includegraphics[clip,scale=0.43]{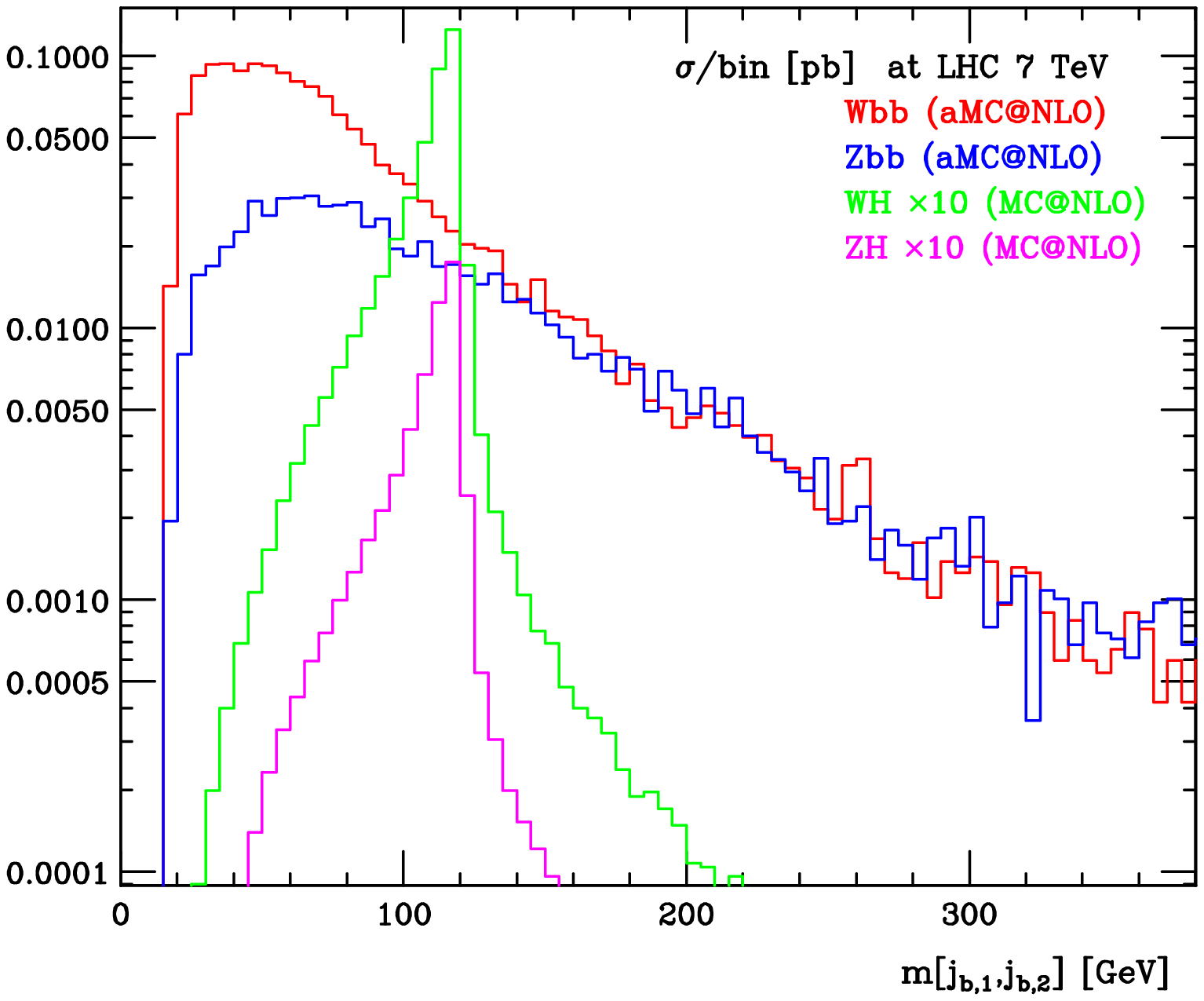}
\hspace{1cm}
\includegraphics[angle=0,scale=0.43]{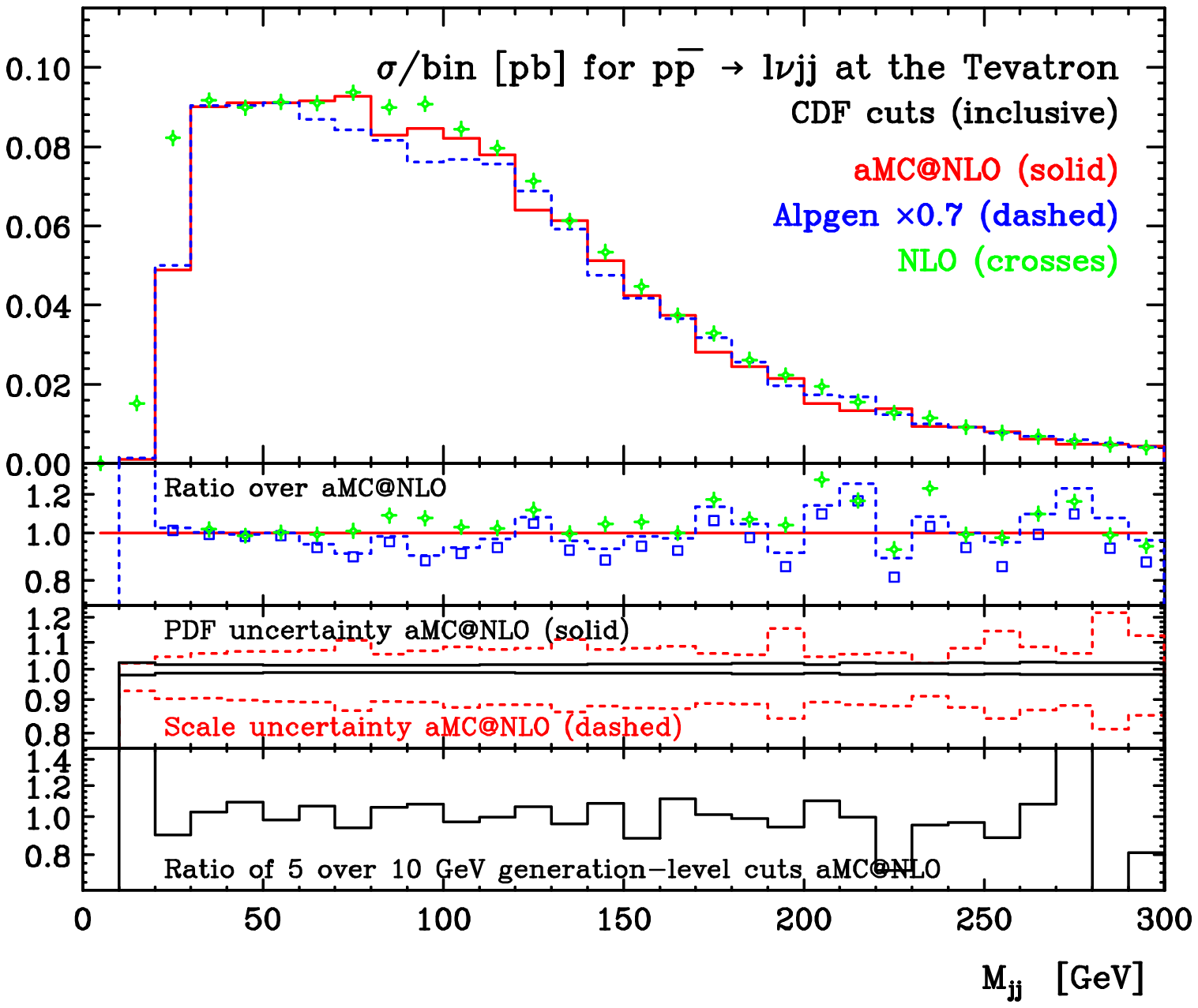}
\caption{Left: Invariant mass of the pair of the two leading $b$-jets
  for $ W b \bar b$, $ Z b \bar b$, $WH(\to \ell\nu b\bar{b})$,
  $ZH(\to\ell^+\ell^- b\bar{b})$ at the LHC (7 TeV), the latter two
  are rescaled by a factor of ten. Figure taken
  from~\cite{Frederix:2011qg}.  Right: Invariant mass of the pair of
  the two hardest jets in $W$+ jet events with CDF/D0 inclusive
  cuts. Figure taken from ref.~\cite{Frederix:2011ig}.}
\label{fig:mbbmjj}
\end{figure}

Similarly, {\tt aMC@NLO}{} is a novel approach to a complete event
generation at NLO.
{\tt aMC@NLO}{} has been used recently for the calculation of $W/Z b
\bar b$~\cite{Frederix:2011qg} and $W$+dijet
production~\cite{Frederix:2011ig}.  Fig.~\ref{fig:mbbmjj} (left) shows
an application to Higgs searches of the $W/Z b \bar b$ calculation:
the invariant mass of the pair of the two leading $b$-jets, for the
processes $Wbb$, $Zbb$, $WH$, and $ZH$.  The figure illustrates a case
where signals and irreducible backgrounds are computed with the same
accuracy. The process $Wb\bar b$ has been implemented shortly before
also in the \POWHEGBOX\cite{Oleari:2011ey}.

Fig.~\ref{fig:mbbmjj} (right) illustrates predictions from {\tt
  aMC@NLO} for the invariant mass for the dijet system in $Wjj$ (the
observable mentioned in the introduction where CDF observed a large
deviation from the SM). CDF and D0 estimate the $Wjj$ using a leading
order Monte Carlo (LO+PS) re-weighted to the NLO cross-section or to
data.  With {\tt aMC@NLO} instead it is possible to compute directly
the $Wjj$ cross-section at the NLO+PS level. It was therefore
particularly interesting to check whether there is any shape
difference between LO+PS and NLO+PS in the $M_{jj}$ distribution. The
study of ref.~\cite{Frederix:2011ig} shows that there is no sizable
shape difference. Another interesting application of {\tt aMC\@NLO}{}
is the calculation of scalar and pseudo-scalar Higgs production in
association with a $t \bar t$ pair~\cite{Frederix:2011zi}.

\begin{figure}
\centering 
\includegraphics[angle=0,scale=0.55]{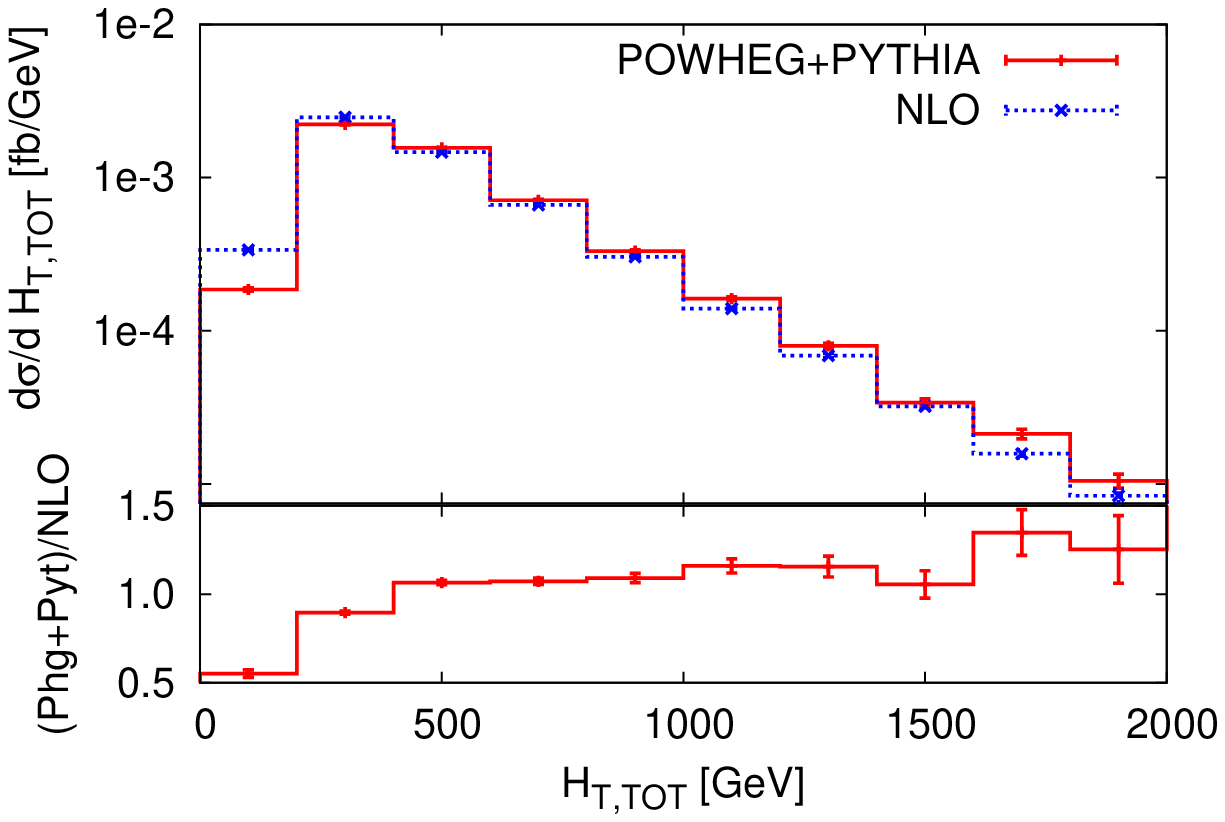}
\includegraphics[angle=0,scale=0.55]{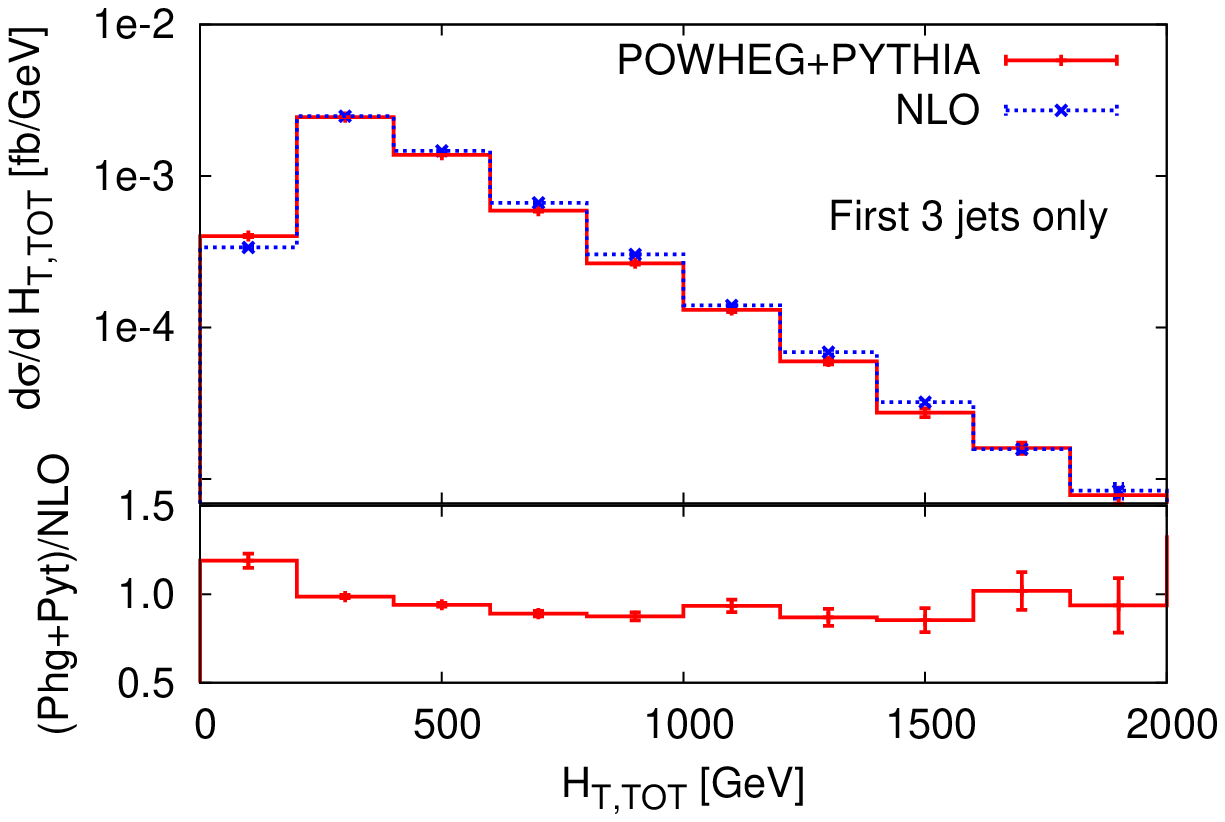}
\caption{Comparison of NLO and \POWHEGpPYTHIA{} results for the
  $H_{\rm T,TOT}$ distribution in the process $W^+W^++$ 2 jets at the
  LHC (7 TeV), when all jets are included in the definition of $H_{\rm
    T,TOT}$ (left pane), and when only the three hardest jets are
  included (right pane). Figure taken from ref.~\cite{Melia:2011gk}.}
\label{fig:httot}
\end{figure}

A lot of effort has been devoted recently also to the implementation
of higher multiplicity processes in the \POWHEGBOX{} and in {\tt
  aMC@NLO}. 
The first $2\to 4$ process that has been implemented in the
\POWHEGBOX{} is $pp \to W^+W^+$+ 2 jets including both the QCD induced
part~\cite{Melia:2011gk} as well as the VBF
contributions~\cite{Jager:2011ms}.  This is a relatively simple $2 \to
4$ process since the cross-section is finite without any cut on the
jets. As expected, for inclusive observables there are only minor
differences between pure NLO and \POWHEG{}+{\tt PS}, but for exclusive
observables, depending on the details of the observable definition,
there can be important differences. This is shown in
Fig.~\ref{fig:httot} for two different definitions of $H_{\rm T, TOT}
=\sum_j p_{\rm t,j}$, the transverse energy of the event. From the
figure it is clear that if only the three hardest jets are included in
the definition of $H_T$, the corrections from the PS are very moderate
(right pane), but if all soft jets present in the event are included,
then additional radiation from the PS can alter the distribution
substantially (left pane).
The \POWHEG{} method has been used recently also for the description
of other important processes, e.g. for the calculation of $Z$ boson in
association with a top anti-top pair at NLO accuracy including PS
effects~\cite{Garzelli:2011is}.

\subsection{\MENLOPS{} and \LoopSim}
In NLO+PS approaches only contributions with one additional jet,
relative to the Born contribution, are computed accurately, while all
other emissions are described only in the shower approximation.  There
are however situations in which one has important contributions from
higher multiplicity final states (e.g. because of new channels that
open up that are enhanced by kinematical factors, gluon PDFs etc.).
\MENLOPS~\cite{Hamilton:2010wh,Alioli:2011nr} is a method to further
improve on NLO+PS predictions with matrix elements involving more
partons in the final state. For example, for $W$ production it
includes, as in \MCatNLO{} or \POWHEG{}, $W$ production at NLO, the
PS, but also $W+1,2,3, \ldots$ jets using exact matrix
elements. Roughly speaking, it uses a jet-algorithm to define two
different regimes, and then corrects the 1-jet fraction using exact
matrix elements and the 2-jet fraction using the NLO $K$-factor. This
achieves NLO quality accuracy for inclusive quantities but an improved
sensitivity to hard radiation and multi-parton kinematic features.

\begin{figure}[t]
  \centering
  \includegraphics[width=0.95\textwidth]{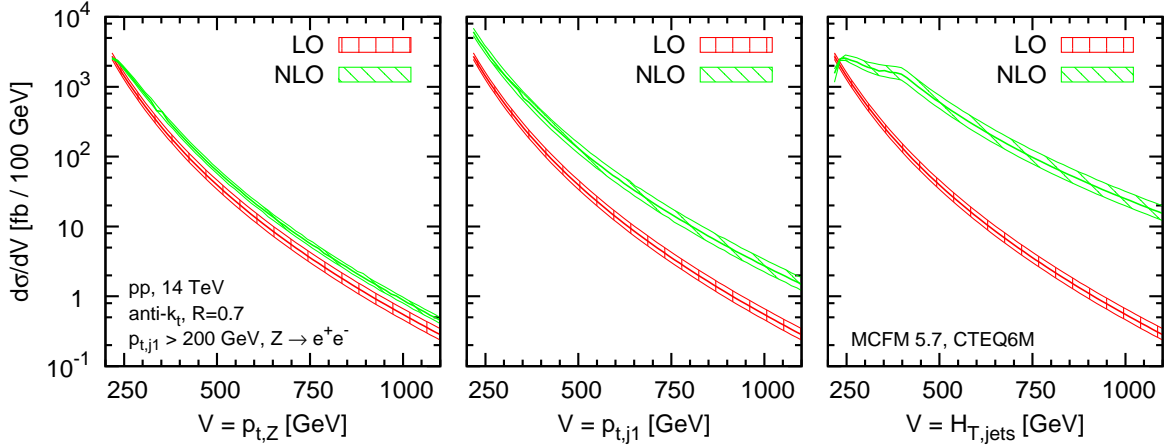}
  \caption{The LO and NLO distributions for three observables in $Z$+jet
    production that are identical at LO: the $Z$ transverse momentum
    (left), the $p_t$ of the hardest jet (middle), and the scalar sum
    of the transverse momenta of all the jets, $H_{\rm T, jets}$
    (right).  The bands correspond to the uncertainty from a
    simultaneous variation of $\mu=\mu_{R}=\mu_F$ by a factor of two
    around the default $\mu=\sqrt{\smash[b]{p_{t,j1}^2 +
        m_Z^2}}$. Figure taken from~\cite{Rubin:2010xp}.}
  \label{fig:Zj:LOandNLO}
\end{figure}

A further recent theoretical development is \LoopSim.  If one
considers the process $W+1$ jet, the three observables $p_{\rm t,Z}$,
$p_{\rm t,j}$, and $H_{\rm T,{\rm jets}} = \sum_j p_{\rm t,j}$ are
identical at LO. However, as illustrated in
Fig.~\ref{fig:Zj:LOandNLO}, at NLO $p_{\rm t,Z}$ has a moderate
$K$-factor ($\lessapprox 2)$, $p_{\rm t,j}$ has a large $K$-factor
($\sim 5$) and $H_{\rm T,jets}$ has a giant $K$-factor ($\sim 50$).
The very large $K$-factors in the last two observables are due to the
fact that the NLO result is dominated by configurations where there
are two hard jets and a soft $W$ (these are enhanced by EW
logarithms), additionally there is an important enhancement coming
from incoming $qq$ channels. \LoopSim~\cite{Rubin:2010xp} is a
procedure that uses a sequential algorithm, close to the
Cambridge/Aachen one, to determine the branching history, ``loops''
over soft particles (i.e. they are removed from the event and the
residual event is adjusted), and it uses a unitary operator to cancel
divergences. In essence, this is a way to extend a calculation that is
exact at a given order in perturbation theory, in an approximate way
to higher orders. The procedure is expected to be more accurate the
larger the corresponding $K$-factor is.
One might expect other extensions of the MLM/CKKW matching procedure
along the same lines as \MENLOPS{} and \LoopSim{} in the near future.

\section{Top-quark production}

The top is the most interesting SM quark. Its large mass implies a
large Yukawa coupling, which causes the top to be a prominent decay
product in many BSM models.  LHC data have already been successfully
compared to approximate next-to-next-to-leading order (NNLO)
predictions~\cite{Aad:2011yb,Chatrchyan:2011yy}, however various
approximate NNLO predictions, based on a threshold resummation, do not
fully agree within quoted
uncertainties~\cite{Kiyo:2008bv,Ahrens:2011mw,Beneke:2011mq,Kidonakis:2011tg,Cacciari:2011hy}.
Significant improvements in the $t\bar t$ cross-sections can be
expected only upon inclusion of the complete NNLO corrections.
A better perturbative control of the top-quark pair production
cross-section is also important to further constrain gluon PDFs, to
have an accurate extraction of the top mass from the cross-section,
and to improve our perturbative control over the $t \bar t$
forward-backward asymmetry.
In fact, an almost $3 \sigma$ deviation from the SM is observed by
CDF, which becomes a $4.2\sigma$ effect in the high-mass region,
$M_{t\bar t}> 450$ GeV~\cite{Aaltonen:2011kc}. The large inclusive
asymmetry has been seen both by CDF and D0~\cite{Abazov:2011rq}, while
the rise in the spectrum of the asymmetry is not confirmed by D0.  One
has however to bear in mind that $t \bar t$ production is a difficult
measurement given the presence of neutrinos in the final state, the
combinatorics in the reconstruction of the tops, and the limited
statistics at the Tevatron. Nevertheless, various suggestions have
been made recently to explain the asymmetry in terms of BSM physics,
but all proposals face the problem that they have to preserve the good
agreement with the symmetric $t \bar t$ observables, respect dijet
bounds and/or must evade the stringent limits on like-sign top
production.
Fervid activity is therefore currently devoted towards a complete NNLO
calculation of $t\bar t$ production (see~\cite{Czakon:2011ve} and
references therein).
Recently, the program {\tt Top++} was released that evaluates
numerically the total inclusive cross-section for producing top quark
pairs at hadron colliders. It calculates the cross-section at fixed
order through approximate NNLO and it includes a soft-gluon
resummation in Mellin space at next-to-next-to-leading logarithmic
accuracy~\cite{Czakon:2011xx}.
The program {\tt Hathor}~\cite{Aliev:2010zk} also calculates the total
cross-section for top-quark pair production in hadronic collisions. It
includes approximate next-to-next-to-leading order perturbative QCD
corrections.  It also offers the possibility to obtain the
cross-section as a function of the running top-quark mass.
A direct comparison between the two programs is however not possible
as they use a different subleading terms in the threshold expansion.

Besides the inclusive cross-section, top pair production in
association with light jets, heavy partons, photons or heavy vector
bosons is also interesting. We know now at NLO $t \bar t$ + 2
jets~\cite{Bevilacqua:2011hy}, $t\bar t b\bar
b$~\cite{Bredenstein:2010rs,Bevilacqua:2009zn}, $t \bar t
Z$~\cite{Lazopoulos:2008de}, $t \bar t
+\gamma$~\cite{Melnikov:2011ta}, and for $t \bar t + 1$ jet including
hard jet radiation by the top quark decay
products~\cite{Melnikov:2011qx}. 
A matching to parton shower has been recently achieved also to $t\bar
t $ + 1 jet production~\cite{Kardos:2011qa,Alioli:2011as}. 

The NLO calculation of single top in the $s$ and $t$
channel~\cite{Frixione:2005vw,Alioli:2009je} and in the $Wt$
channel~\cite{Frixione:2008yi,Re:2010bp} have also been matched to
parton showers.
For a discussion of recent theoretical progress in single-top physics
at hadron colliders, and of aspects of single-top production in beyond
Standard Model scenarios I refer the reader to
ref.~\cite{Motylinski:2011ab}.

\section{A few theoretical issues in Higgs production}
Possibly the most awaited result during the first year of running of
the LHC concerned Higgs searches. 
Currently, both ATLAS and CMS reached the expected sensitivity, around
or better than the SM cross-section.
ATLAS restricted the most likely mass range at 95\% confidence level
(CL) to the region $115.5-131$ GeV. They observe an excess around
$126$ GeV, with a local significance of 3.6$\sigma$, with
contributions from the main channels $H\to\tau \tau$, $H\to ZZ\to 4
l$, $H\to WW\to 2l2 \nu$. The global significance taking into account
the look-elsewhere-effect is 2.3$\sigma$.
CMS excludes the region $127-600$ GeV at 95\% CL (while
their expected exclusion is $117-543$ GeV). They could not exclude the
region below 127 GeV since data has a modest excess of events between
115 and 127 GeV that appears, quite consistently, in five independent
channels. The excess is compatible with a SM Higgs hypothesis in the
vicinity of 124~GeV or slightly below, but the statistical
significance, 2.6$\sigma$ local and 1.9$\sigma$ global after
correcting for the look-elsewhere-effect in the low mass region, is
not large enough to say anything conclusive. 
For both experiments, what is observed now is consistent either
with a background fluctuation or with a SM Higgs boson. More refined
analyses and additional data in 2012 will definitely give an answer.

At the LHC, the Higgs is mainly produced via an intermediate top loop
in gluon-gluon fusion.  The urge to understand the EW symmetry
breaking led in the past years to the computation of the most advanced
theoretical predictions for this process. For instance we now know the
main $g g \to H$ production mechanism including NLO corrections with
exact top and bottom quarks in the loop~\cite{Spira:1995rr}, NNLO
corrections in the large $m_t$
limit~\cite{Harlander:2002wh,Anastasiou:2002yz,Ravindran:2003um}, EW
corrections~\cite{Actis:2008ug}, mixed QCD-EW
corrections~\cite{Anastasiou:2008tj}, and resummation of large
logarithms possibly with N$^3$LO soft
effects~\cite{Catani:2003zt,Moch:2005ky,Ravindran:2005vv,Ravindran:2006cg,Laenen:2005uz,Ahrens:2010rs}. Threshold
corrections to the boson rapidity distribution are also
known~\cite{Ravindran:2006bu}, these higher order corrections
stabilise the theoretical predictions under scale variations.
Furthermore, the most
advanced codes~\cite{Anastasiou:2005qj,Grazzini:2008tf} allow for
fully exclusive decays of the Higgs to $\gamma \gamma$, $W^+W^-\to
e^+\nu e^-\bar \nu$, and $ZZ\to 4l$. A similar accuracy has been
reached recently also in associated $V H$ production where, because
this process is an important one if the Higgs is light, the decay of
the Higgs into $b\bar b$ has been
considered~\cite{Ferrera:2011bk}. NLO EW corrections to $WH/ZH$
production including the vector-boson decays are also
known~\cite{Denner:2011rn}.
As expected, the EW corrections, which are at the level of (5-10)\%
for total cross sections, increase with increasing transverse momentum
cuts. For instance, for $p_{\rm T,H}>200$~GeV, which is an interesting
range at the LHC, the EW corrections to $WH$ production are of the
order of -15\% for $M_H=120$~GeV.

The fully differential decay of a light Higgs boson to bottom quarks
at NNLO in perturbative QCD has been computed
recently in~\cite{Anastasiou:2011qx}. From a technical point of view,
it is interesting to note that this work constitutes the first
physical application of a novel method of non-linear mappings for the
treatment of singularities in the radiative processes which contribute
to the decay width.
The program {\tt iHixs}~\cite{Anastasiou:2011qw} has also been
recently released, it computes the inclusive Higgs boson
cross-section, including QCD corrections through NNLO, EW corrections,
mixed QCD-electroweak corrections~\cite{Anastasiou:2008tj}, quark-mass
effects through NLO in QCD, and finite width effects for the Higgs
boson and the heavy quarks. Furthermore, it allows the evaluation of
the cross-section in modified Higgs boson sectors with anomalous
Yukawa and EW interactions as appearing in some extensions of the
SM~\cite{Anastasiou:2010bt,Anastasiou:2011qw}.

Given the high accuracy with which gluon-gluon-fusion has been
computed, it is interesting to ask what is the actual theoretical
uncertainty on this process. Unfortunately, there is today no full
consensus on this question. Some more conservative estimates quote
errors of the order of 40\%~\cite{Baglio:2010um} (at the Tevatron) and
similar uncertainties at the LHC, while other studies suggest that the
perturbative uncertainty is considerably smaller.
Assigning a correct theoretical error is very important when claiming
an exclusion or an excess, and, at a later stage, when making
measurements of the Higgs-boson couplings, which is the only way to
identify the precise nature of the Higgs boson and EW symmetry
breaking. Yet, even for the main Higgs-production channel there are
still some controversies and subtleties. Most controversies have to do
with how different sources of errors should be combined, others
concern the question of how to assign/interpret the perturbative
uncertainties. I will illustrate here just two of these issues.

The soft logarithms appearing in cross-sections can be resummed using
an effective theory approach. Performing such a calculation requires
an introduction of a matching scale, where the full and effective
theory amplitudes must agree. It is well-known that choosing a
time-like (i.e. complex) matching scale effectively resums $\pi^2$
enhanced terms. In~\cite{Ahrens:2008nc} it is suggested that this
procedure improves the convergence of the perturbative expansion
significantly, and reduces the uncertainty of the perturbative (NNLO)
prediction. This approach is criticized in ~\cite{Dittmaier:2011ti}
with the arguments that $\pi^2$ are just numbers, so that there is no
formal limit in which they dominate, and that only one class of
$\pi^2$ terms is resummed (those that arise from the gluon form
factor), but not all of them. In this context, one has to mention that
perturbative QCD is often about pushing approximations beyond their
formal limit of validity, and that a given approach should be judged
by seeing how well it fares in practice.

\begin{figure}
  \centering
  \includegraphics[width=0.5\textwidth]{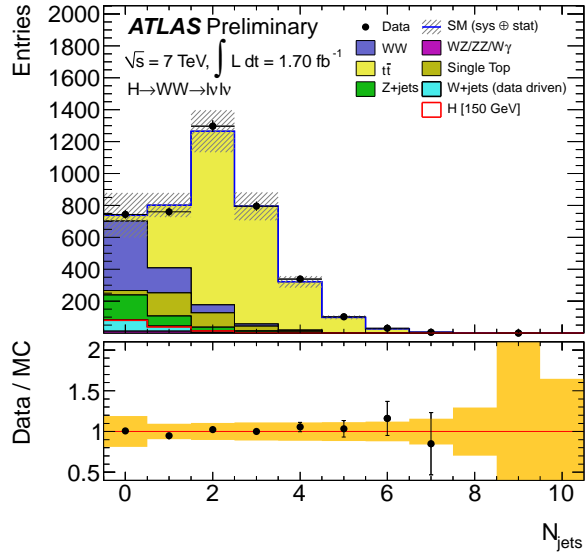}
  \caption{Multiplicity of jets with $p_{\rm T}>25$ GeV after basic Higgs
    search cuts. The lower part shows the ratio between the data and
    the background expectation from MC, with the yellow band
    indicating the total systematic uncertainty in the normalization
    (but not the shape) of the various components. The signal is shown
    for $m_H=150$~GeV. Figure taken from~\cite{ATLAStopveto}.}
  \label{fig:topveto}
\end{figure}

The second issue I would like to mention here has to do with a
jet-veto in Higgs searches. As can be seen from
Fig.~\ref{fig:topveto}, in Higgs searches one needs to impose a
jet-veto to suppress the large top background.  Higgs production is
then studied in 0-, 1-, 2-jet bins separately in order to maximize the
sensitivity. Currently, ATLAS uses $p_{\rm t, veto}= 25$ GeV, while
CMS employs $p_{\rm t, veto} = 30$ GeV in their Higgs searches.

\begin{figure*}[t!]
\includegraphics[width=0.5\textwidth]{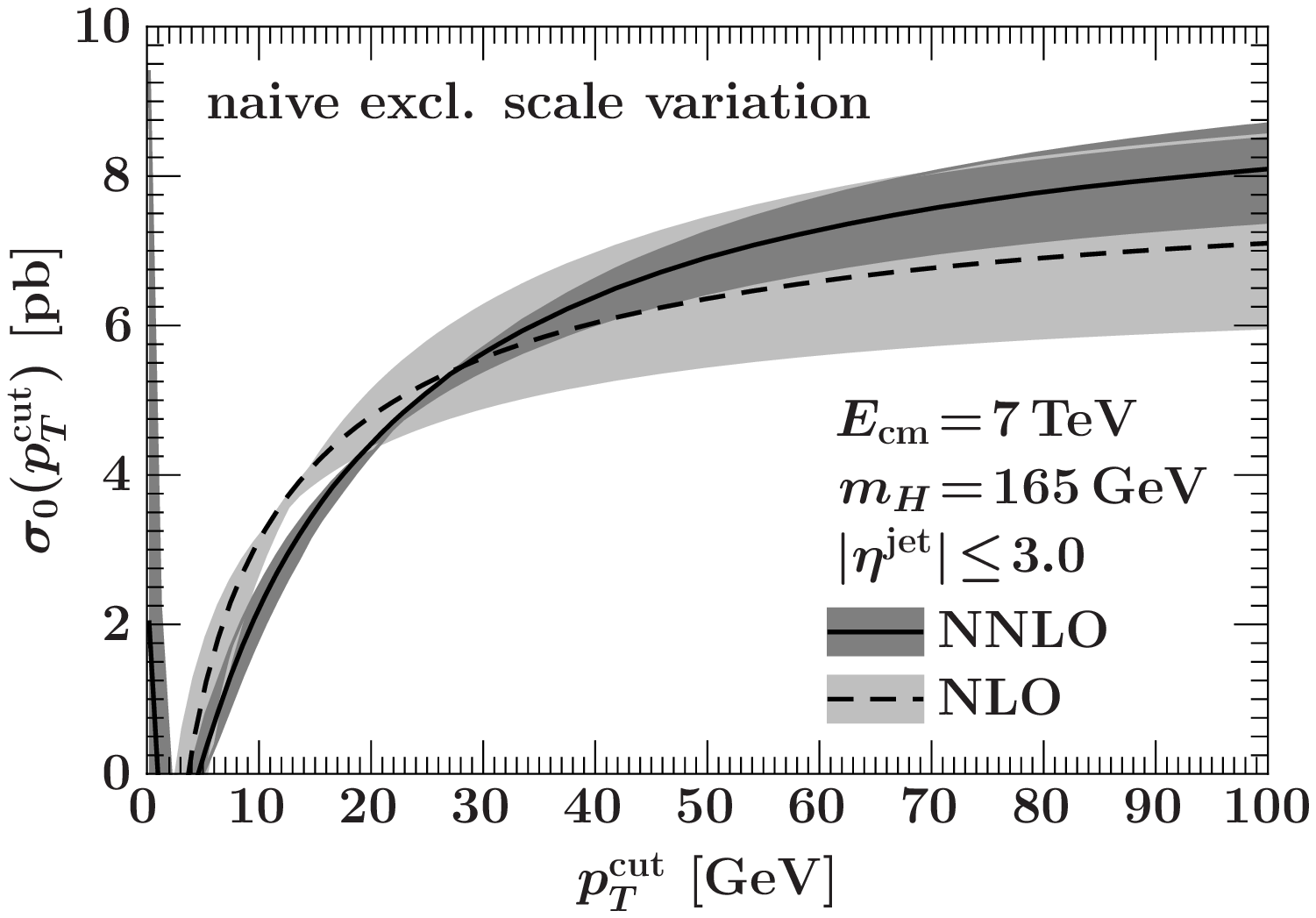}%
\hfill%
\includegraphics[width=0.5\textwidth]{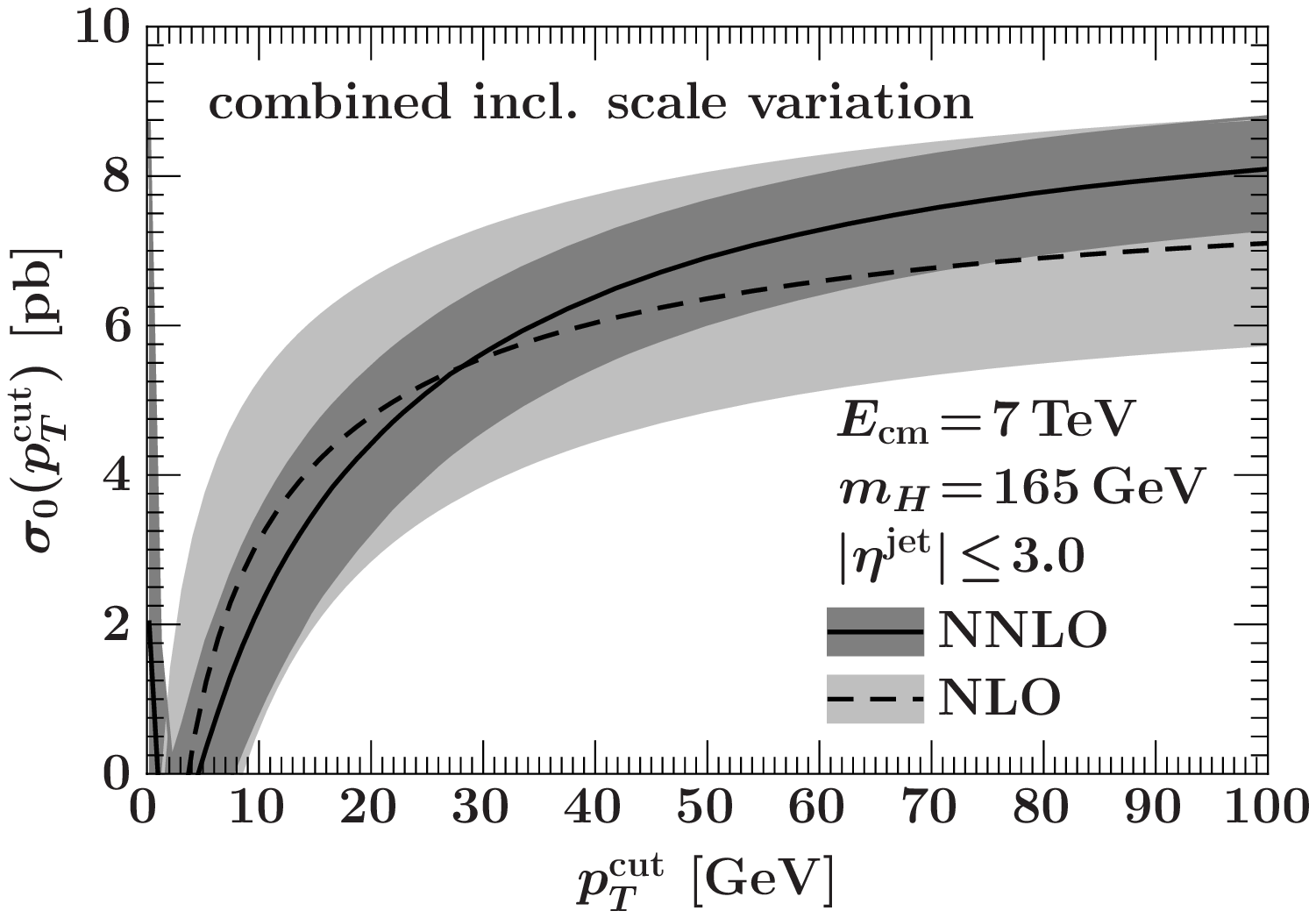}%
\caption{Fixed-order perturbative uncertainties for $gg\to H + 0$
  jets at NLO and NNLO for the LHC (7 TeV). On the left, the
  uncertainties are obtained from the naive scale variation in
  $\sigma_0(p_{\rm T}^{\rm cut})$ between $\mu = m_H/4$ and $\mu =
  m_H$. On the right, the uncertainties are obtained by independently
  evaluating the scale uncertainties in $\sigma_{\rm tot}$ and
  $\sigma_{\geq 1}(p_{\rm T}^{\rm cut})$ and combining them in
  quadrature. Figure taken from~\cite{Stewart:2011cf}.}
\label{fig:jetveto} 
\end{figure*}

In~\cite{Anastasiou:2009bt} the inclusive, NNLO Higgs production
cross-section at the Tevatron is split into 0-,1-jet exclusive, and
2-jet inclusive components
\begin{equation}
\frac{d\sigma_{\rm tot}}{\sigma_{\rm tot}} = 66.5\%^{+5\%}_{-9\%}
(0{\rm -jet})
+28.6\%^{+24\%}_{-22\%} (1{\rm -jet})
+4.9\%^{+78\%}_{-41\%} (\ge 2 {\rm -jets})
= [-14.3\%+14.0\%]\,.
\label{eq:sighiggs}
\end{equation}
The errors denote the scale uncertainty that is obtained by
varying the renormalization and the factorization scale together
around a central value $m_{\rm H}=160$ GeV by a factor of two.
In an NNLO calculation of inclusive Higgs production, only the 0-jet
bin is known at NNLO, while the 1-jet bin is know at NLO and the
2-jet bin is computed at LO only. Therefore is it not surprising
that the relative errors increase with the number of jets.
In~\cite{Anastasiou:2009bt} one can also find a detailed discussion of
why it is {\it not} appropriate to use the standard scale variation as
an estimate of the perturbative uncertainty for the 0-jet bin
cross-section (Fig.1 of~\cite{Anastasiou:2008ik} also shows that for a
particular choice of $p_{\rm T,veto}$ one obtains a vanishing scale
uncertainty band in the 0-jet bin).
The numbers in eq.~(\ref{eq:sighiggs}) were updated by Campbell {\it
  et al.} in~\cite{Campbell:2010cz} who evaluated the 2-jet bin
contribution at NLO. The effect of this addition was a slight change
in all relative numbers, and, mainly, a decrease in the perturbative
uncertainty of the 2-jet bin,
\begin{equation}
\frac{d\sigma_{\rm tot}}{\sigma_{\rm tot}} = 60\%^{+5\%}_{-9\%}
(0-{\rm jet})
+29\%^{+24\%}_{-23\%} (1-{\rm jet})
+11\%^{+35\%}_{-31\%} (\ge 2 {\rm jets})
= [-15.5\%+13.8\%]\,.
\label{eq:sighiggs2}
\end{equation}
From eq.~(\ref{eq:sighiggs2}), it is evident that the scale
uncertainty is smaller for the {\it exclusive} measurement with
0-jets, than the one of the fully {\it inclusive} measurement.  To
explain this feature, Stewart and Tackmann recall that there are two
mechanisms at work in the 0-jet cross-section~\cite{Stewart:2011cf}:
there is a large $K$-factor from perturbative higher orders, as well
as large negative logarithms $-\alpha_s C_A/\pi\ln^2M_H/p_{\rm t,
  veto}$ that become more important the smaller $p_{\rm t, veto}$
is. They therefore suggest that the error on the 0-jet bin should be
computed taking into account a full correlation between jet-bins,
i.e. the error from the 0-jet cross-section is computed from the
relation $\sigma_0 = \sigma_{\rm incl} - \sigma_{\ge {\rm
    1-jet}}$. One obtains then simply $\Delta^2 \sigma_0 =
\Delta^2\sigma_{\rm incl} + \Delta^2\sigma_{\ge {\rm 1-jet}}$. The
effect of this is illustrated in Fig.~\ref{fig:jetveto}.  While this
procedure is certainly more conservative than a conventional scale
variation, it is clear that to reduce the uncertainty on the jet-veto
cross-section, a resummation of large logarithms involving the ratio
$p_{\rm t, veto}/M_H$ is required. Currently, only resummation for
quantities related to the jet-veto exist, e.g. for $p_{\rm T,
  Higgs}$~\cite{Bozzi:2005wk} or for the
beam-thrust~\cite{Berger:2010xi}. Both observables are however not the
ones used in current Higgs searches. Furthermore the beam thrust has
the drawback that it receives very large non-perturbative corrections,
as can be easily seen by running a PS program once at parton and once
at hadron level.

\section{Gauge boson production processes}

The physics programme involving gauge bosons is particularly rich.
The road towards precision measurements and searches starts from the
measurement of inclusive $W$ and $Z$ cross-sections, which, from a
theoretical point of view, are the most precisely known processes at
hadron colliders.  Beyond measurements of purely inclusive $W/Z$
production cross-sections, it is possible to study the production of
$W$ or $Z$ bosons in association with one or more jets, and ratio of
cross-sections. Interesting ratios are for instance
$\sigma(V+(n+1)\,{\rm jets})/\sigma(V+n\,{\rm jets})$, with $V=W,Z$,
that start at ${\cal O}(\alpha_s)$ in the perturbative expansion in
the coupling constant, or ratios of $\sigma(W^{\pm}+n\, {\rm
  jets})/\sigma(Z+n\,{\rm jets})$, that are of order ${\cal
  O}(\alpha_s^0)$. In both cases, one expects many experimental and
theoretical uncertainties (e.g. those related to the choice of
renormalization or factorization scale, or uncertainties in the parton
distribution functions) to largely cancel in the ratio. A further
extension of simple ratios are asymmetry distributions, for instance
the $W$ or lepton charge asymmetry. These distributions provide strong
constraints on parton distribution functions and are useful probes of
new physics.
Other interesting observables measure gauge bosons produced in
association with heavy quarks (charm or bottom quarks).  These
processes are particularly interesting because of the discrepancies
between theoretical predictions and Tevatron
data~\cite{Aaltonen:2009qi}, however the perturbative calculation of
these processes and of the related theoretical uncertainty is
challenging.
Finally, di-boson cross-sections are sensitive to any type of new
physics that would modify the trilinear gauge couplings, and would
give rise to so-called anomalous gauge couplings. These measurements
are complementary to the ongoing direct searches for beyond the SM
physics, and are able to probe new-physics scales that are not
directly accessible. 

\begin{figure}[t]
  \centering
\includegraphics[angle=0,scale=0.6]{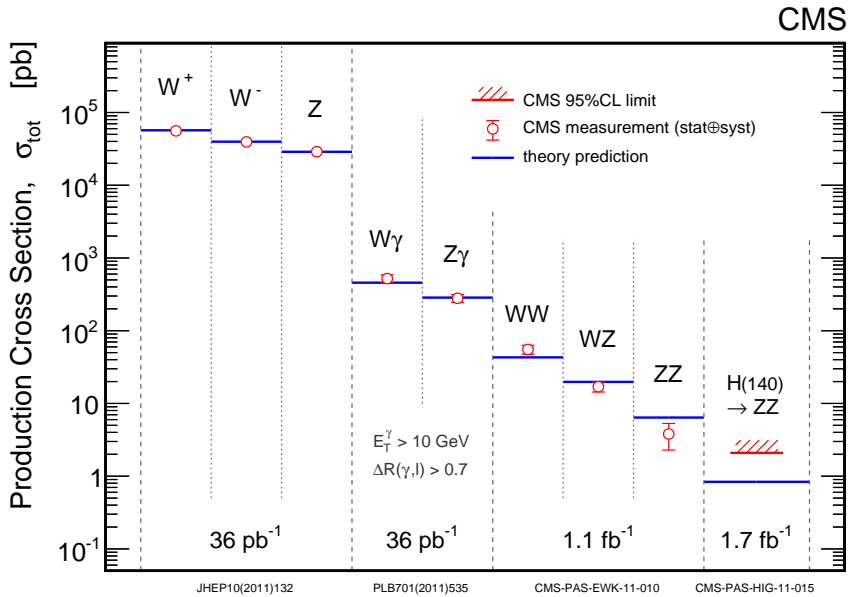}
\caption{
Production cross-sections for processes involving gauge bosons at the
LHC (at $\sqrt{s}=7$ TeV). Figure
taken from~\cite{cmspage}.} 
\label{fig:EW}
\end{figure}

Fig.~\ref{fig:EW} gives the cross-section for main processes involving
gauge bosons at the LHC.  The figure illustrates that with 1fb$^{-1}$
ATLAS and CMS could collect ${\cal O}(10^6)$ and ${\cal O}(10^5)$ $W$
and $Z$ events per experiment and per lepton channel. One also sees
that, including all lepton channels, 1fb$^{-1}$ of data contain about
100 $WW$ and 10 $ZZ$ events. This means that even with the data
available after a first year of running, a number of interesting
analyses could be performed.

\subsection{Drell-Yan}
\begin{figure}[t]
  \centering
  \includegraphics[width=0.65\textwidth]{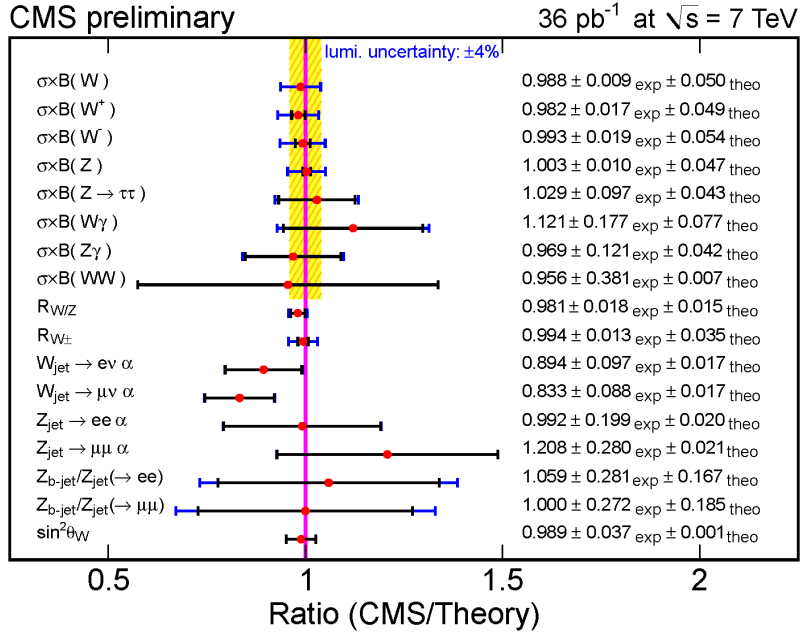}
  \caption{Comparison of NNLO theory and CMS data for Drell-Yan
    observables. Figure taken from ref.~\cite{CMSprelim}.}
  \label{fig:DY}
\end{figure}

The most important gauge boson production process is Drell-Yan. This
is the best known process at the LHC: it has been computed through
NNLO in QCD, fully differential in lepton momenta including
spin-correlations, EW corrections, finite-width effects, and
$\gamma^*/Z$ interference.  State-of the art codes are described
in~\cite{Catani:2009sm,Gavin:2010az}.  Calculations to all-orders also
exist, for instance the next-to-next-to-leading logarithms (NNLL)
transverse momentum resummation~\cite{Bozzi:2010xn} and soft gluon
resummation have been computed~\cite{Cao:2009md}.
Threshold enhanced corrections to rapidity distributions of $Z$ and
$W^\pm$ bosons at hadron colliders are also known at N${}^3$LO
level~\cite{Ravindran:2007sv}.
These accurate perturbative calculations have been available for some
time, and now that precise LHC data has been compared to those
predictions, one can not but praise the impressive agreement between
NNLO theory and experiment (see e.g. Fig.~\ref{fig:DY} and
ref.~\cite{Chatrchyan:2011cm}). One thing to note is that in these
comparisons of theory with experiment, the dominant error is the
theoretical one, however this is mainly dominated by the luminosity
uncertainty (of the order of 4\%).

At this level of precision, it is legitimate to start worrying about
mixed QCD and EW corrections.  QCD+EW interference is formally a
higher-order effect, and exact results are not known
currently. However the dominant effects come from emissions in the
soft/collinear regions where an approximate factorization holds.  It
turns out that in general mixed corrections are small, ${\cal
  O}(0.5\%)$, but they are enhanced in the region of large transverse
momenta, where they can reach up to around
5\%~\cite{Balossini:2009sa}.

\subsection{Charge asymmetry}
\begin{figure}[t]
  \centering
  \includegraphics[width=0.7\textwidth]{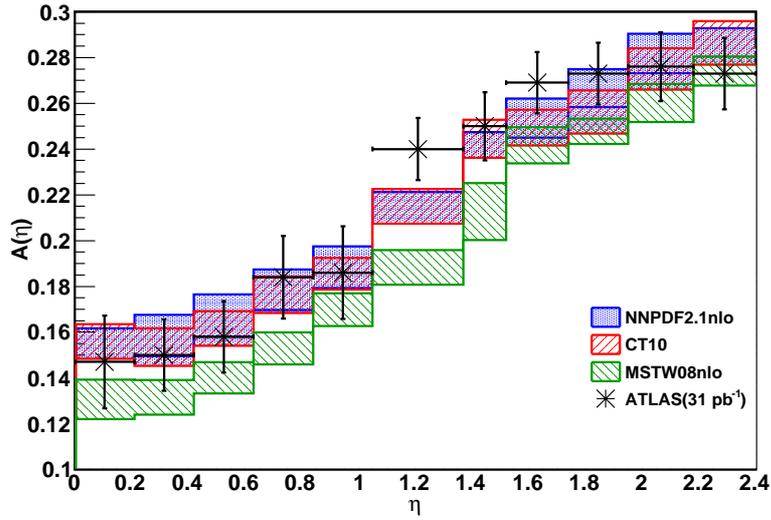}
  \caption{Predictions for the $W$ lepton asymmetry at NLO, obtained
    with DYNNLO~\cite{Catani:2009sm} using the CT10~\cite{Lai:2010vv},
    MSTW08~\cite{Martin:2009iq} and NNPDF2.1~\cite{Ball:2011uy} parton
    sets, compared to measurements for the muon charge asymmetry from
    ATLAS~\cite{Aad:2011yn} (7 TeV). Figure taken from
    ref.~\cite{Ball:2011gg}.}
  \label{fig:LepAsym}
\end{figure}

The natural extension of the inclusive cross-section is the $R_W =
W^+/W^-$ ratio. One can then study $R_W$ as a function of kinematical
variables, e.g. one can look at the charge asymmetry as a function of
lepton rapidity $\eta$
\begin{equation}
A(\eta) = \frac{R_W(\eta)-1}{R_W(\eta)+1}\,.
\end{equation}
This measurement is very sensitive to PDFs since in the ratio
asymmetric properties of PDFs are enhanced, while many uncertainties
cancel. Fig.~\ref{fig:LepAsym} shows the relative good agreement of
theoretical predictions that use various PDFs with ATLAS data. It also
illustrates how the shape of the theoretical prediction is sensitive
to the PDFs chosen.
Indeed, ATLAS and CMS measurements of this distribution have been
already used by the Neural Network (NN) collaboration to constraint
PDFs. In particular a reduction of uncertainty of the
order of $10-30\%$ in the range $x=[10^{-3}, 10^{-1}]$ was obtained
for the valence- and sea-quark distributions.
It is interesting to observe that LHCb data at larger rapidities probe
larger and smaller values of $x$ that are currently less
constraint. They will therefore soon have a larger impact in PDF
determination than ATLAS and CMS have.

\subsection{Production of $W/Z$ boson in association with jets}

\begin{figure}[t]
  \centering
  \includegraphics[width=0.5\textwidth]{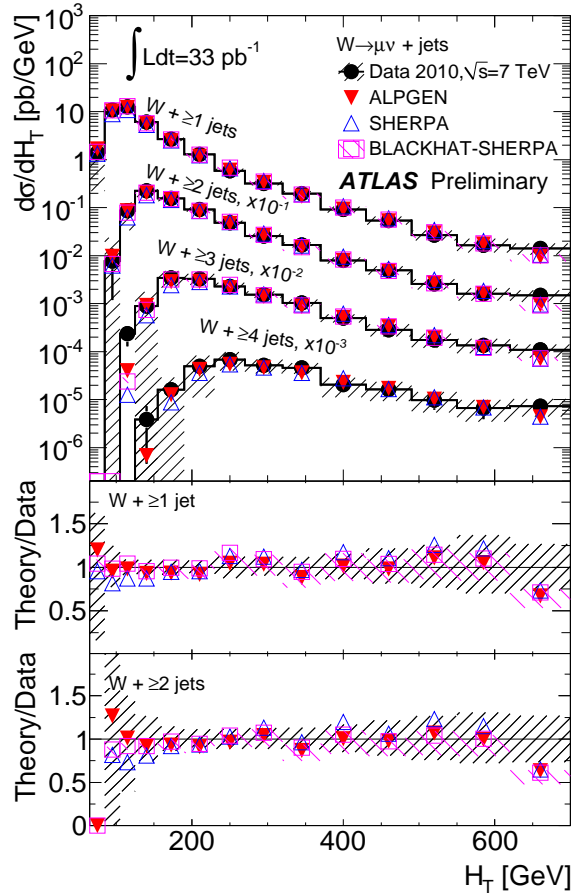}
  \caption{Cross-section for $W\to e \nu$ and jets at the LHC (7 TeV)
    as a function of the total transverse momentum of the event
    $H_{\rm T}$. Figure taken from~\cite{Ask:2011cg}.}
  \label{fig:wplusjets}
\end{figure}

At the LHC, because of the large energy, the production of $W$ and $Z$
bosons in association with jets is very likely. This is illustrated in
Fig.~\ref{fig:wplusjets} which shows the differential distribution for
$H_{\rm T}$, the total transverse energy of the event, for various jet
multiplicities. Since the cross-sections with an additional jet is
rescaled in the figure by a factor $10^{-1}$ compared to the
cross-section with one less jet, it is evident that at high $H_{\rm
  T}$ (a region of particular interest for various new-physics
searches), all jet-multiplicities contribute similar
amounts.\footnote{Of course, this statement depends on the precise
  definition of the jets, in particular on their $p_{\rm t}$ cut.}
Because of this, it becomes very important to have a good perturbative
control of processes involving the production of $W/Z$ bosons together
with many jets. The perturbative calculation of processes involving a
large number of jets is quite difficult beyond LO.  However, as
discussed in Sec.~\ref{sec:NLO}, recent years have seen a revolution
in the techniques used for NLO calculations.
These novel techniques allowed in the last five years the calculation
of a large number of processes involving gauge bosons and jets. In
particular, while $V+1$ and $V+2$ jets have been described to NLO in
QCD since 1983~\cite{Ellis:1981hk,Giele:1993pk} and
2002~\cite{Bern:1997sc,Nagy:1998bb,Campbell:2002tg}, to quote a few
examples we know now at NLO $VV+1$
jet~\cite{Campbell:2007ev,Dittmaier:2007th,Binoth:2009wk,Campanario:2010hp},
$W+3$ jets~\cite{KeithEllis:2009bu,Berger:2009zg}, $Z+3$
jets~\cite{Berger:2010vm}, $W^+W^+$ plus dijets~\cite{Melia:2010bm} ,
$W^+W^-$ plus dijets~\cite{Melia:2011dw} $W^+W^-
bb$~\cite{Bevilacqua:2010qb,Denner:2010jp} and $W+4$
jets~\cite{Berger:2010zx} and $Z+4$ jets~\cite{Ita:2011wn}.
Furthermore, various vector boson fusion (VBF) induced gauge boson
production processes have been computed and are available in the
public code \VBFNLO~\cite{Arnold:2011wj}.

\begin{figure}[t]
  \centering
  \includegraphics[width=1.0\textwidth]{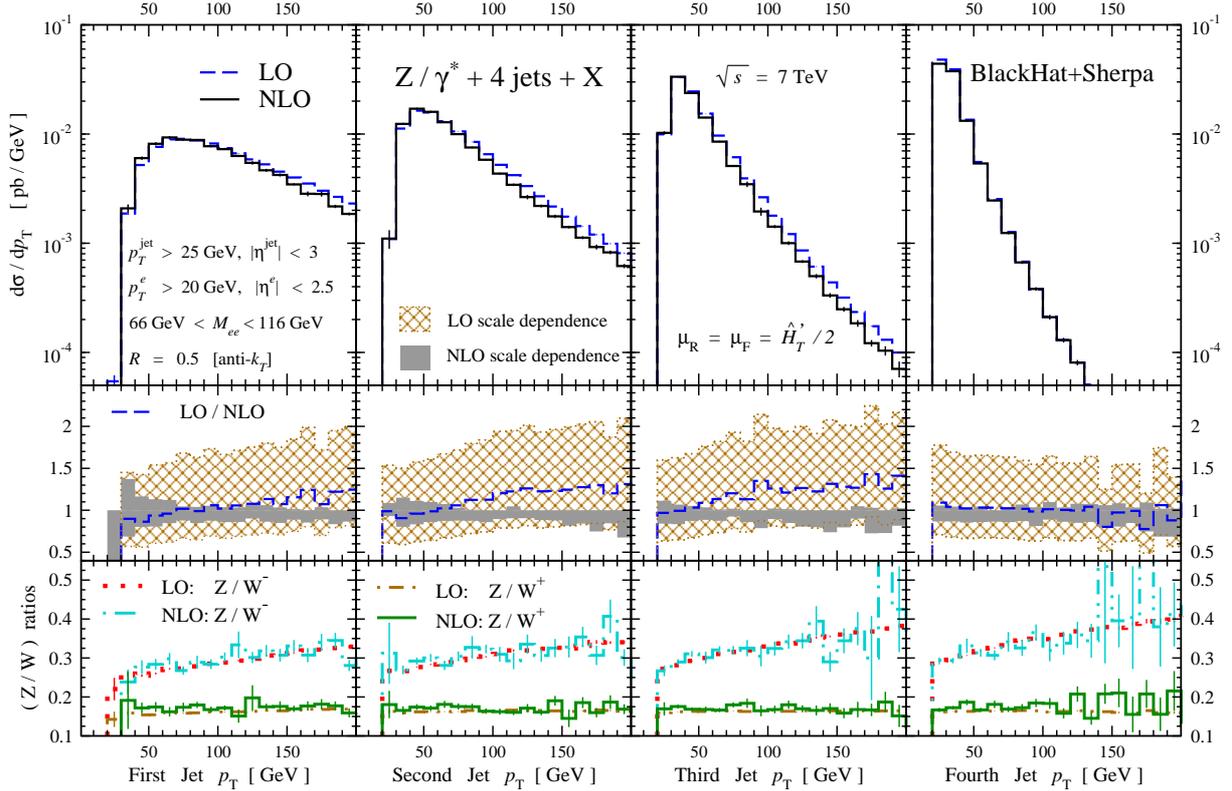}
  \caption{Transverse momentum distributions of the leading four jets
    in $Z+4$-jet production at the LHC (7 TeV). Figure taken
    from~\cite{Ita:2011wn}.}
  \label{fig:Z4j}
\end{figure}

Fig.~\ref{fig:Z4j} shows the transverse momentum distribution for the
four $p_{\rm t}$ ordered jets in $Z+4$ jets production. The middle
pane illustrates the typical reduction of the dependence of the
cross-section on renormalization and factorization scale at NLO,
compared to LO. It is however also evident that while for some
distributions (e.g. $p_{\rm t,j4}$) the ratio of LO/NLO is flat, in
other cases the shape of the NLO distribution is different from the LO
one, so that LO/NLO has a non-flat slope. Is it also interesting to
look at the ratio of $Z/W^+$ and $Z/W^-$ displayed in the lower
pane. The agreement of LO and NLO predictions for these ratios
illustrates the excellent perturbative control that one can achieve at
NLO on these ratios.  Furthermore, the fact the $Z/W^+$ is flat, while
$Z/W^-$ rises with the $p_{\rm t}$ can be attributed to the fact that
both $W^+$ and $Z$ production are dominated by the $u$-quark
distribution, while $W^-$ is mainly produced from $d$ quarks in the
protons, therefore the $Z/W^-$ ratio displays the slope of the
$u/d$-quark distributions as a function of the jet transverse momenta.

\subsection{Di-boson production and anomalous couplings}

The $ZZ$, $WW$, $WZ$ di-boson production of processes that have been
implemented recently in the \POWHEGBOX. The calculation includes
$\gamma^*/Z$ interference, single resonant contributions, interference
effects for identical fermions and, for $WW$ and $WZ$ the effect of
anomalous couplings. The gluon-gluon fusion contribution, that is
formally NNLO, but is important when Higgs search cuts are applied, is
available the \ggtoWW{} and \ggtoZZ{}
generators\cite{Binoth:2008pr,Binoth:2005ua,Binoth:2006mf}, and,
recently, also from the program \MCFM~\cite{Campbell:2011cu}.

A pure NLO calculation, as implemented earlier in
\MCFM~\cite{Campbell:1999ah,Campbell:2011bn}, reveals that for these
processes the conventional scale variation of the LO result is very
modest but underestimates completely the size of the NLO
corrections. E.g. for $pp\to W^+W^-\to e^+\nu_e\mu^-\nu_\mu$
production at the 7 TeV LHC without any cuts, the LO cross-section
using NNPDf2.1~\cite{Ball:2011uy} is $375.2^{+1.6}_{-3.8}$ fb, while
the NLO cross-section is $499.8^{+12}_{-10}$
fb~\cite{Melia:2011tj}. Here the error denotes the scale uncertainty.
Similar results (with smaller cross-sections) are obtained for the
other processes. The reason for the large NLO corrections (not caught
by the LO scale variation) is that new partonic channels open up at
NLO. It is therefore clear that only a NLO calculation can provide a
reliable estimate of the cross-section and of its error. These
di-boson production processes are particularly interesting since they
are important backgrounds to Higgs searches. Furthermore, they are
sensitive to new physics at high scales through the measurement of
anomalous trilinear couplings (ATGCs).
Indeed, while the LHC does probe new physics at the TeV scale
directly, ATGCs indirectly probe physics in the multi-TeV range, since
they arise when high-energy degrees of freedom are integrated
out. Both the Tevatron \cite{:2009us,Abazov:2009tr,Abazov:2009ys} and
LEP \cite{Alcaraz:2006mx} were able to place quite stringent bounds on
ATGCs.  However, since their effects are enhanced at high energies,
one expects even better bounds from the LHC. Indeed, CMS already
presented bounds on the anomalous couplings appearing in an effective
Lagrangian with the parametrization of ref.~\cite{Hagiwara:1993ck}
without form factors \cite{Chatrchyan:2011tz}.

\begin{figure}[t]
\centering 
\includegraphics[width=0.6\textwidth]{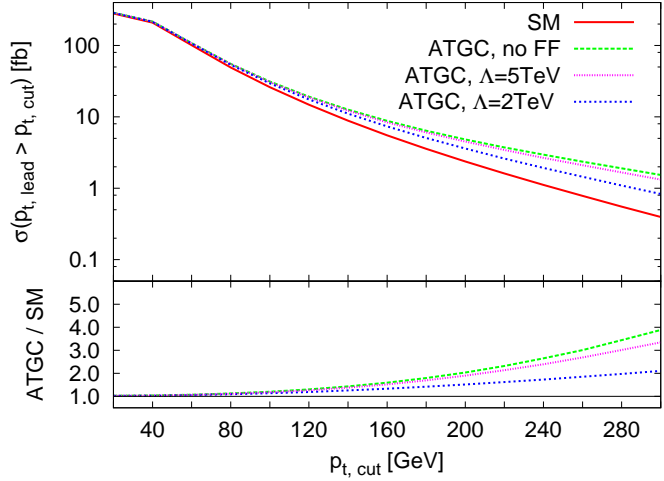}
\caption{The integrated cross-section for $W^+W^-$ production at the
  LHC (7 TeV) as a function of the cut on the transverse momentum of
  the leading lepton is shown on the right. Figure taken
  from~\cite{Melia:2011tj}. }
\label{fig:ptlead} 
\end{figure}

Following refs.~\cite{Hagiwara:1986vm,Baur:1987mt,Hagiwara:1993ck},
one can parametrize the most general terms for the $WWV$ vertex
($V=\gamma,Z$) in a Lagrangian that conserves $C$ and $P$ as
\newcommand\gwwv{g_{{\scriptscriptstyle WWV}}}
\newcommand\gwwz{g_{{\scriptscriptstyle WWZ}}}
\newcommand\sW{{\scriptscriptstyle W}}
\newcommand\sV{{\scriptscriptstyle V}}
\newcommand\gwwg{g_{{\scriptscriptstyle WW}\gamma}}
\begin{equation}
\begin{split}
{\cal L}_{\rm eff} = i \gwwv & \Big[ g_1^V 
(W^*_{\mu\nu} W^\mu V^\nu - W_{\mu\nu}W^{*\mu} V^\nu)\\
&.+\kappa^V W^*_{\mu}W_{\nu} V^{\mu\nu}
+\frac{\lambda^V}{M_\sW^2}W^*_{\mu\nu}W^{\nu}_{\rho} V^{\rho\mu}
\Big]\,,
\end{split}
\label{leff}
\end{equation}
where $W_{\mu\nu} = \partial_\mu W_\nu - \partial_\nu W_\mu$, $\gwwz=
-e\, {\rm cot} \theta_\sW$ and $\gwwg = -e$.  In the SM $g_1^V =
\kappa_1^V =1$, and $\lambda^V= 0$. Any departure from these values
($\Delta g_1^V = g_1^V-1$ etc.) would be a sign of new physics. In the
\POWHEG{} generator, all six parameters can be set independently.

In the presence of anomalous couplings, the effective Lagrangian of
eq.~\eqref{leff} gives rise to interactions that violate unitarity at
high energy. Thus, in order to achieve a more realistic
parametrization, the couplings are multiplied by form factors, that
embody the effects arising from integrating out the new degrees of
freedom. The precise details of the form factors therefore depend on
the particular model considered. Paralleling the discussion of
ref.~\cite{Baur:1988qt}, in the \POWHEGBOX{} it is assumed that all
anomalous coupling $\Delta g$ are modified as
\begin{equation} 
\Delta g \to \frac{\Delta g}{(1+ M_{\sV\sV}^2/\Lambda^2)^2}\,,
\label{ff}
\end{equation}
where $M_{\sV\sV}$ is the invariant mass of the vector boson pair and
$\Lambda$ is the scale of new physics.

Fig.~\ref{fig:ptlead} sets the anomalous coupling to the maximum
deviation from the SM allowed by LEP bounds and displays the
sensitivity of the transverse momentum of the leading jet in $W^+W^-$
events to anomalous couplings, for the case of a form factor
$\Lambda=2$ TeV, $\Lambda=5$ TeV, and no form factor
($\Lambda=\infty$).  The plot illustrates the great potential of the
LHC to improve on existing bounds (even more so, when the machine will
run at yet higher energy).

For the production of four charged leptons, a similar NLO study (again
including all off-shell, spin-correlation, virtual-photon-exchange,
and interference effects) can be performed with the {\tt aMC@NLO}{}
generator~\cite{arXiv:1110.4738}.  The parton showering can be done
either with {\tt HERWIG}{} and (for the first time in this context)
with {\tt Pythia 6}{}.  The $O(\alpha_s^2)$ contribution of the $gg$
channel is also included directly here, as obtained from {\tt
  MadLoop}~\cite{Hirschi:2011pa}. In the study of
ref.~\cite{arXiv:1110.4738} several key distributions together with
the corresponding theoretical uncertainties are presented. A further
theoretical improvement is that scale and PDF uncertainties are
computed at essentially no extra CPU-time using re-weighting
techniques.

\section{Parton distribution functions and $\alpha_s$}

PDFs are the second thing one can not live without, if one works on
LHC physics.  Huge effort is devoted today in understanding
differences and improving the theoretical and statistical treatment of
PDFs. This activity is reflected in new PDFs sets being released by
various groups~\cite{Alekhin:2011sk}.
The main focus of all groups is now directed towards NNLO PDFs sets,
an improvement in the treatment of heavy quarks, an introduction of
flexible parametrization, a more dynamic tolerance, and, of course,
towards the inclusion of more data in the fits. Discussions are
ongoing that try to clarify whether discrepancies between different
PDFs are due to the inclusion of different data sets. For instance,
there is no full consensus on what impact of the Tevatron jet data has
on gluon distributions at the LHC.

\begin{figure}[t]
\includegraphics[angle=0,scale=0.41]{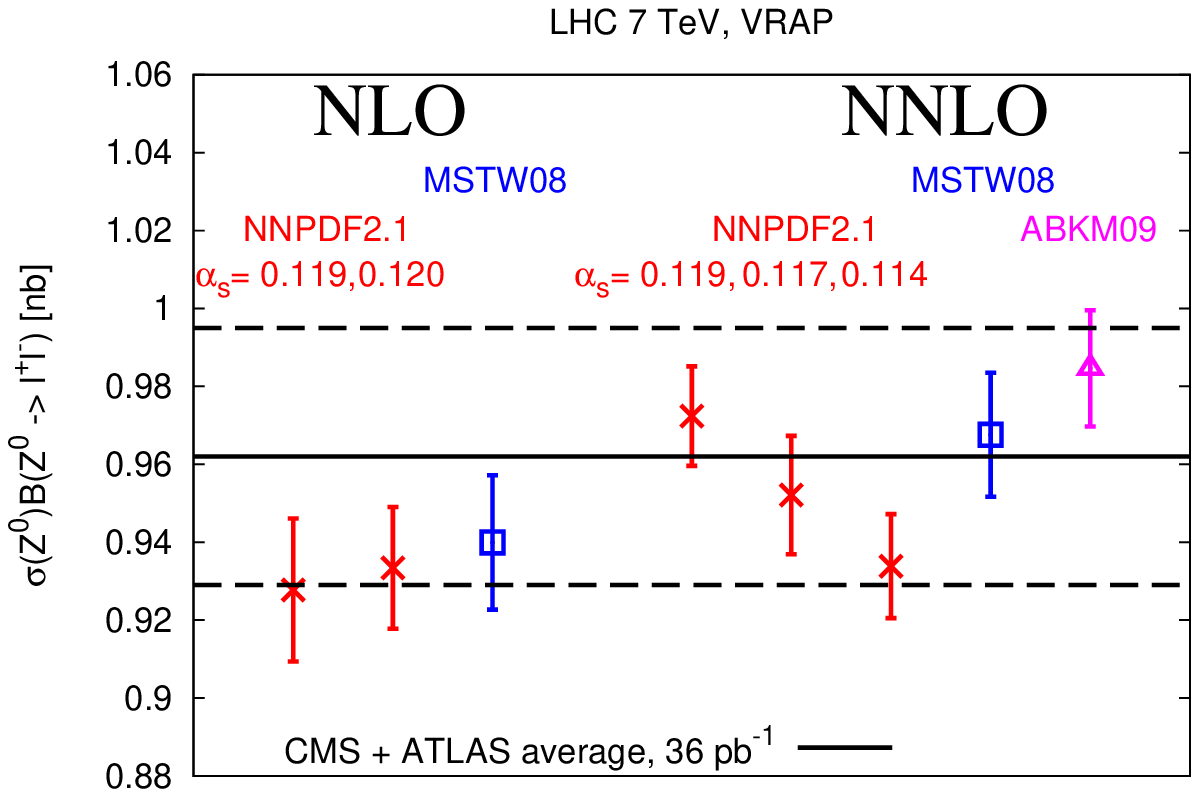}
\includegraphics[angle=0,scale=0.41]{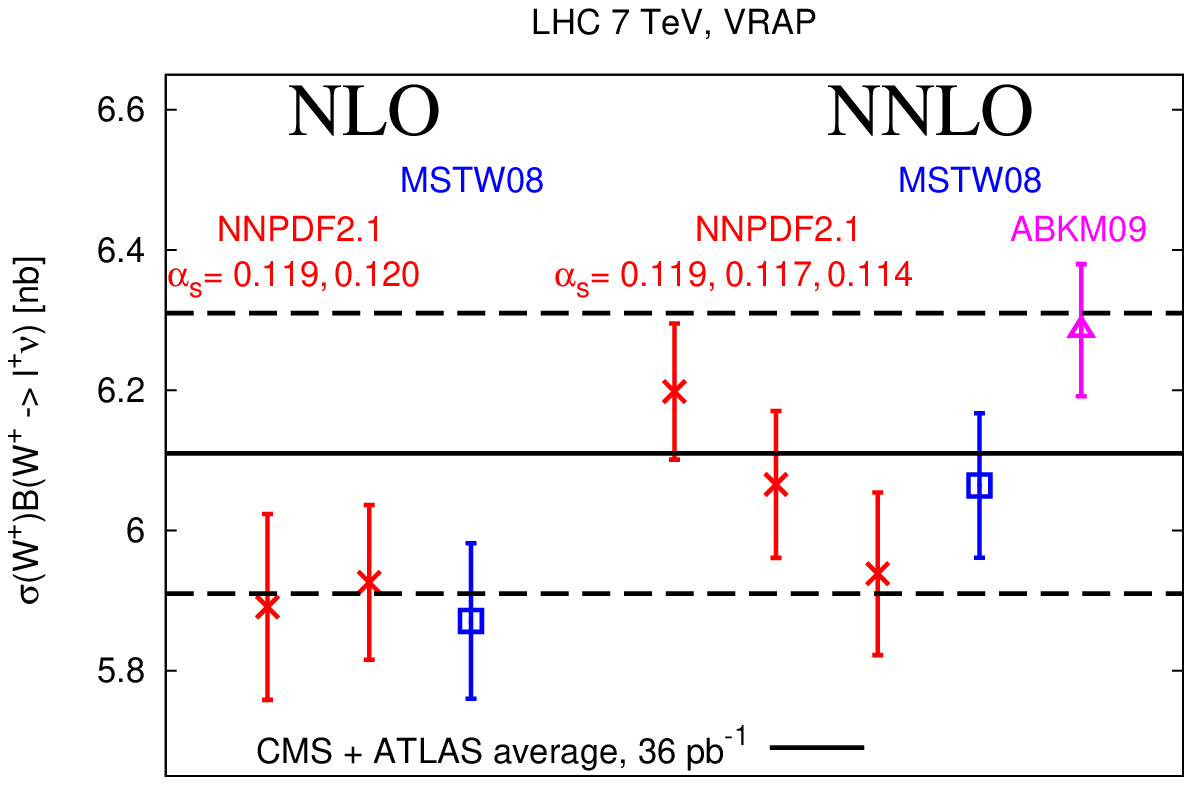}
\includegraphics[angle=0,scale=0.41]{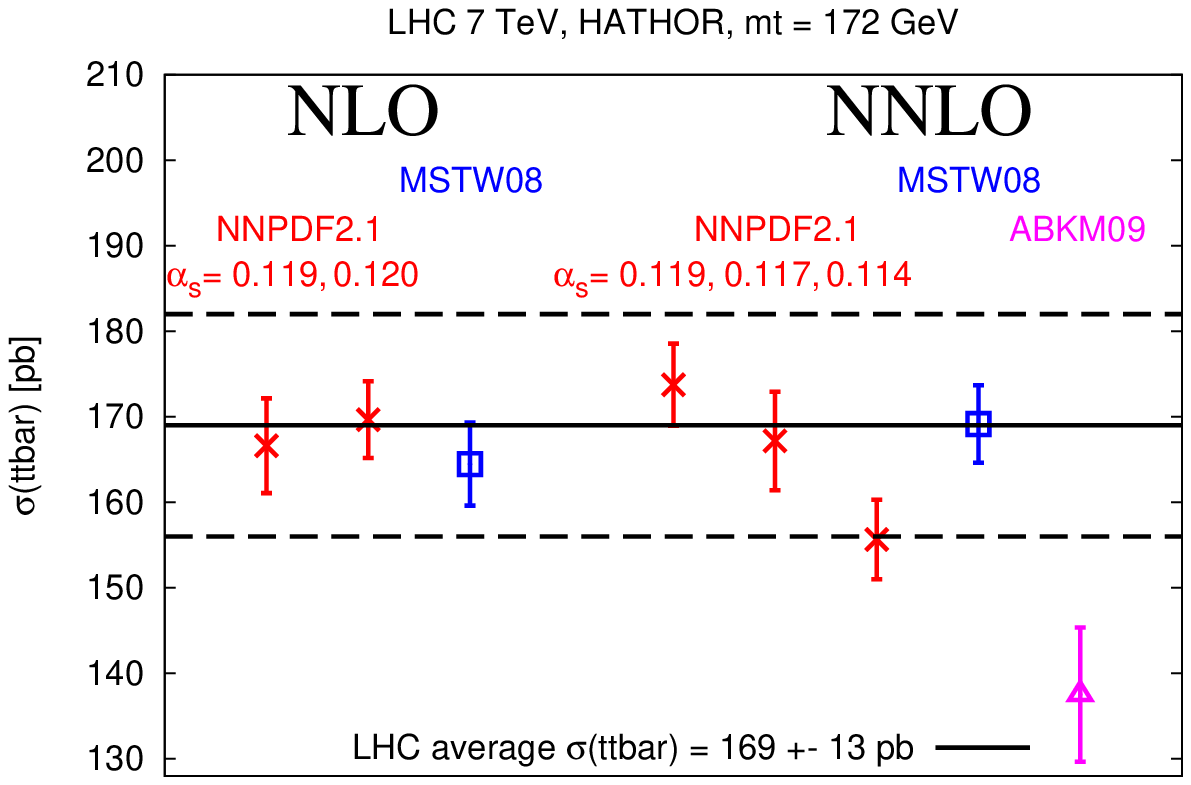}
\caption{The total cross-section for $Z$, $W^+$, and $t\bar t$, at the
  LHC (7 TeV) for NNPDF2.1 with $\alpha_s(M_Z)=0.119$ (NLO and NNLO),
  $\alpha_s(M_Z)=0.120$ (NLO) and $\alpha_s(M_Z)=0.114,0.117$ (NNLO),
  MSTW08 with $\alpha_s(M_Z)=0.1202$ (NLO) and $\alpha_s(M_Z)=0.1171$
  (NNLO), and ABKM09 with $\alpha_s(M_Z)=0.1135\pm0.0014$ (NNLO).
  Uncertainties shown correspond to $1\sigma$.  The band corresponds
  to the combination of CMS and ATLAS measurements. Figure taken from
  ref.~\cite{Ball:2011uy}.}
\label{fig:tt-wp-z}
\end{figure}

Fig.~\ref{fig:tt-wp-z} shows the uncertainty on three LHC benchmark
processes ($Z$, $W^+$, and $t\bar t$ from the left to the right)
coming from using different PDFs or a different value of $\alpha_s$,
at NLO and at NNLO. Differences are due to the inclusion of different
data in the fits, due to a different methodology (e.g. the
parametrization), due to a different treatment of heavy quarks, and
due to a different default value of the coupling constant. In
particular, it is remarkable how much benchmark processes depend on
the value of $\alpha_s$. The preliminary 2011 average value of
$\alpha_s$ is $\alpha_s = 0.1183\pm 0.0010$~\cite{Bethke:2011tr}. It
is interesting to note that the value barely changed compared to the
2009 number ($\alpha_s = 0.1184 \pm 0.0007$)~\cite{Bethke:2009jm}, but
that the uncertainty on it increased. This is due to the inclusion of
new data in the fits which tend to move the average value in opposite
directions. An open issue today, in the combination of the various
measurements to produce a world average for $\alpha_s$, is the
treatment of outliers that have very small errors. This is the case
for the extraction of $\alpha_s$ from thrust computed at N$^3$LL
including power corrections using SCET~\cite{Abbate:2010xh}, for the
number obtained from $\tau$-decays in~\cite{Pich:2010xb}, and for the
hadronic event shapes in $e^+e^-$ collisions at OPAL using NNLO+NLLA
theoretical predictions~\cite{opal:2011qm}, just to quote the most
important cases.

New processes added to the world average since 2009 include inclusive
jets at the Tevatron~\cite{Abazov:2009nc}, the $e^+e^-$ 3-jet rate
which is know to NNLO~\cite{Dissertori:2009qa}, and $e^+e^-\to $ 5
jets, which is now known at NLO~\cite{Frederix:2010ne}. The use of the
Tevatron jet data in this context is particularly interesting: while
the error on $\alpha_s$ from this extraction is not particularly
small, this measurement and the sensitivity of benchmark processes on
the value of $\alpha_s$ shown in Fig.~\ref{fig:tt-wp-z} raises the
question of whether it is possible to make competitive measurements of
$\alpha_s$ at the LHC. The extraction of the value of the coupling
constant at hadron colliders must take into account that PDFs
themselves do depend on $\alpha_s$. A viable possibility then is to
consider appropriate ratios (e.g. $W/(Z+(n+1) \, {\rm jets})/(Z/(W+n
\, {\rm jets}))$).

\section{Jet algorithms}
Jet algorithms are the third thing you can not live without, if you
work on LHC physics.  For a long time, infrared (IR) unsafe algorithms
were used at the Tevatron, with several ``patches'' introduced to
minimize the effect of the IR-unsafety.  At the LHC, both ATLAS and
CMS have adopted as default the anti-$k_t$
algorithm~\cite{Cacciari:2008gp}. Given that this algorithm was
proposed only three years ago, it shows how flexible experimentalists
are today in adopting new, successful ideas.\footnote{A minor downside
  to this is that ATLAS and CMS use a different radius -- the choices
  for ATLAS are 0.4 and 0.6, while for CMS they are 0.5 and 0.7.}
Using this algorithm both collaborations have already explored scales
up to $4$ TeV and could place constraints on various BSM models, in
particular those models that would give rise to a resonance in the
$M_{jj}$ distribution (such as massive coloured bosons, black-holes,
\ldots).

\begin{figure*}[t!]
\centering
\includegraphics[width=0.5\textwidth]{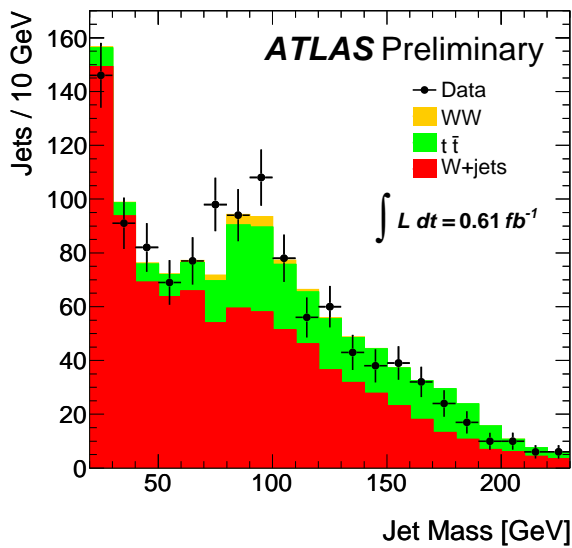}%
\caption{Jet mass in $W$ + jets for a jet with substructure compatible
  with $WZ(\to bb)$ or $WH(\to bb)$ events. Figure taken from
  ref.~\cite{EPSboostedhiggs}.}
\label{fig:hbb} 
\end{figure*}

Other IR-safe algorithms like the Cambridge-Aachen or SISCone are in
use as well. These are particularly useful for studies which exploit
the fact that when a massive boosted object decays, it gives rise to a
``fat jet'' with a non-trivial jet-substructure. Looking at the
internal structure of these jets using jet-grooming techniques like
filtering, pruning or trimming has a huge potential for making
discoveries
``easier''~\cite{Abdesselam:2010pt,Altheimer:2012mn}. These techniques
have a big gain in sensitivity over traditional methods, but one might
lose many events when imposing strict kinematical cuts and requiring a
boosted regime. The potential of these studies has been demonstrated
in several examples~\cite{Abdesselam:2010pt,Altheimer:2012mn}. However
sophisticated jet studies are still a young field, and as of now there
are no precise rules on how to make discoveries easier. What is
impressive, is that even these very new techniques are already being
used at the LHC.
Fig.~\ref{fig:hbb} shows for instance the single hadronic jet mass in
$W$+ jet events in a boosted regime, an observable relevant for $W
H(\to bb)$. In Fig.~\ref{fig:hbb} the $Z$ peak coming from $WZ(\to
bb)$ is evident, these very first results seem therefore very
promising in finding a possible peak to due Higgs production.

\section{Conclusions}
The physics programme at the LHC is very rich: it spans from most
precise measurements, e.g. in the case of Drell-Yan, to searches with
highest reach for new physics either direct searches, or indirect ones
(e.g. through the potential presence of anomalous couplings).
This experimental programme at the LHC is supplemented by robust
theoretical predictions that include NLO QCD corrections, mixed
NLO-QCD+EW corrections, NNLO, and resummation of logarithmically
enhanced contributions. Furthermore, different calculations are merged
using clever matching procedures that catch the best features of
different calculations.
From the theoretical community, there is a clear and successful effort
to produce predictions and public codes that have the flexibility
required for today's sophisticated experimental analysis
(e.g. including parton shower, decays with spin correlations, massive
quark effects, etc.). 

Impressive results have already come out of the LHC, but this is
certainly only the tip of the iceberg.  After just one year of running
at the LHC some measurements, e.g. in processes involving $W/Z$
production, start being dominated by theoretical and parton density
errors.  It will therefore be a real challenge for theorists to keep
up with the high experimental precision.

\end{document}